\documentclass[a4paper,11pt]{article}

\usepackage{jheppub} 
\usepackage{graphicx,color}
\usepackage{amsmath}
\usepackage{slashed}
\usepackage{multirow}
\usepackage{autobreak}
\usepackage[normalem]{ulem}
\usepackage{hyperref}
\usepackage{cleveref}
\usepackage{slashed}
\usepackage{amssymb}
\usepackage{epsfig}
\usepackage{array}
\usepackage{mathtools}
\usepackage{caption}
\usepackage[utf8]{inputenc}
\usepackage{mathtools, nccmath}
\usepackage{float}
\usepackage[section]{placeins}
\usepackage[utf8]{inputenc}
\usepackage[T1]{fontenc}
\usepackage{lmodern}
\usepackage[x11names]{xcolor}
\usepackage{framed}
\usepackage[utf8]{inputenc}
\usepackage[dutch]{babel}
\colorlet{shadecolor}{blue!10}

\SetSymbolFont{largesymbols}{bold}{OMX}{txex}{b}{n}

\DeclareMathAlphabet{\mathpzc}{OT1}{pzc}{m}{it}

\newcommand{\be}{\begin{eqnarray*}}
\newcommand{\ee}{\end{eqnarray*}}

\usepackage{parskip}

\newcommand{\ba}{\begin{array}}
\newcommand{\ea}{\end{array}}
\newcommand{\bd}{\begin{displaymath}}
\newcommand{\ed}{\end{displaymath}}
\newcommand{\besub}{\begin{subequations}}
\newcommand{\eesub}{\end{subequations}}

\def\q2 {q^2}

\def\bt{\begin{table}}
\def\et{\end{table}}

\usepackage{bbm}  
\usepackage{amsmath}                                                

\newcommand{\nc}{\newcommand}
\nc{\beq}{\begin{equation}}  \nc{\eeq}{\end{equation}}
\nc{\bea}{\begin{eqnarray}}  \nc{\eea}{\end{eqnarray}}
\nc{\baa}{\begin{array}}     \nc{\eaa}{\end{array}}
\nc{\bit}{\begin{itemize}}   \nc{\eit}{\end{itemize}}
\nc{\ben}{\begin{enumerate}} \nc{\een}{\end{enumerate}}
\nc{\bce}{\begin{center}}    \nc{\ece}{\end{center}}
\nc{\bpm}{\begin{pmatrix}}   \nc{\epm}{\end{pmatrix}}
\nc{\bvt}{\begin{verbatim}}  \nc{\evt}{\end{verbatim}}
\nc{\bal}{\begin{align}}
\def\to{\rightarrow}

\def\boldoverdot{\,{\raise6pt\hbox{\bf.}\!\!\!\!\>}}

\def\ee{{\bf e}}

\usepackage{bm}
\def\diag{\hbox{\diag}}

\def\vevof#1{\left\langle #1 \right\rangle}

\def\doubleundertext#1{
{\undertext{\vphantom{y}#1}}\par\nobreak\vskip-\the\baselineskip\vskip4pt%
\undertext{\hbox to 2in{}}}
\def\inbox#1{\vbox{\hrule\hbox{\vrule\kern5pt
     \vbox{\kern5pt#1\kern5pt}\kern5pt\vrule}\hrule}}
\def\sqr#1#2{{\vcenter{\hrule height.#2pt
      \hbox{\vrule width.#2pt height#1pt \kern#1pt
         \vrule width.#2pt}
      \hrule height.#2pt}}}

\def\today{\ifcase\month\or
  January\or February\or March\or April\or May\or June\or
  July\or August\or September\or October\or November\or December\fi
  \space\number\day, \number\year}
\def\pmb#1{\setbox0=\hbox{#1}%
  \kern-.025em\copy0\kern-\wd0
  \kern.05em\copy0\kern-\wd0
  \kern-.025em\raise.0433em\box0 }

\def\pmbb#1{\setbox0=\hbox{#1}%
  \kern-.02em\copy0\kern-\wd0
  \kern.04em\copy0\kern-\wd0
  \kern-.02em\raise.03464em\box0 }
\def\sumprime_#1{\setbox0=\hbox{$\scriptstyle{#1}$}
  \setbox2=\hbox{$\displaystyle{\sum}$}
  \setbox4=\hbox{${}'\mathsurround=0pt$}
  \dimen0=.5\wd0 \advance\dimen0 by-.5\wd2
  \ifdim\dimen0>0pt
  \ifdim\dimen0>\wd4 \kern\wd4 \else\kern\dimen0\fi\fi
\mathop{{\sum}'}_{\kern-\wd4 #1}}
\DeclareUnicodeCharacter{2212}{-}

	\title{\boldmath Optimal estimation of Dimension-8 Neutral Triple Gauge Couplings at $e^+ e^-$ Colliders}

\author{Sahabub Jahedi\footnote{Corresponding author.}}

\affiliation{Department of Physics, Indian Institute of Technology, Guwahati, Assam 781039, India}

\emailAdd{sahabub@iitg.ac.in}

\abstract{We investigate the measurement of non-standard $ZZV \, (V=\gamma, Z)$ couplings through $Z$-boson pair production at the $e^+ \, e^-$ colliders. We adopt the Standard Model Effective Field Theory (SMEFT) approach to study these anomalous neutral triple gauge couplings. There are one CP-conserving and three CP-violating dim-8 SMEFT operators that contribute to $ZZV$ couplings. Using the optimal observable technique, the sensitivity of these new physics couplings has been estimated and compared with the latest experimental limits on dim-8 couplings at the LHC at CERN. The effect of beam polarization and correlations among CP-violating $ZZV$ couplings are discussed. The comparison of statistical limits of new physics couplings between optimal observable technique and contemporary cut-based analysis has also been studied in detail.}


\begin{document}

\maketitle
\flushbottom

\section{Introduction}
\label{sec:intro}
Gauge boson self interactions within the $SU(2)_L \times U(1)_Y$ gauge symmetry of the Standard Model (SM) provides a unique window to test the SM experimentally and yields an interesting way to look for possible new physics (NP) beyond SM (BSM). The self interactions of charged gauge boson ($W^\pm$) with the neutral gauge bosons ($\gamma,Z$) appear in SM and have been tested with a significant precision at LEP2 and LHC \cite{L3:2004ulv,CMS:2013ant,CMS:2019ppl}. As there is no self interaction among the neutral gauge bosons within SM, neutral triple gauge couplings (nTGCs) ($ZZV, \, V=\gamma,Z$) play a crucial role to search for possible BSM physics. As no NP signal beyond SM is found in the experiment, it is anticipated that the new physics (NP) signature will be very small. The SMEFT framework \cite{Buchmuller:1985jz,Grzadkowski:2010es,Lehman:2014jma,Bhattacharya:2015vja,Liao:2016qyd,Murphy:2020rsh,Li:2020gnx} is an accomplished framework to estimate a small NP signal such as nTGCs in a model independent way. In this framework, nTGCs emerge via gauge-invariant dim-8 effective operators at tree level. We will stick to the approach followed by some earlier works \cite{Hagiwara:1986vm, Renard:1981es, Gounaris:1983zn, Baur:1992cd, Gounaris:1996rz} where the most general non-standard vertex functions are written in terms of anomalous couplings. Thereafter, these anomalous couplings are parameterized in terms of dim-8 effective couplings. 

In order to estimate nTGCs, $Z$-boson pair production at the $e^+e^-$ colliders has gained substantial recognition in the past. Within the SM, the process $e^+e^- \to ZZ$  is governed by $t$-channel electron exchange. As there are no $ZZ\gamma$ or $ZZZ$ couplings in the SM, $s$-channel $ZZ$ production is strictly forbidden at tree-level. One loop $s$-channel mediation is also highly suppressed\footnote{One loop analysis of $e^+e^- \to ZZ$ has been done in \cite{Demirci:2022lmr}.}. In consequence, any deviation from the SM prediction of this process will serve as a useful hint toward an NP signal. In literature, $ZZ$ production has been explored at $e^+e^-$ colliders \cite{Gounaris:1999kf,Rahaman:2016pqj,Rahaman:2017qql,Cetinkaya:2023uip} and $pp$ colliders \cite{Baur:2000ae,Sun:2012zze,Rahaman:2018ujg,Yilmaz:2019cue,Yilmaz:2021ule} in the context of precision measurement of nTGCs\footnote{For other works related to nTGCs, see \cite{Choudhury:1994nt,Atag:2004cn,Ots:2004hk,Ots:2006dv,Ananthanarayan:2011fr,Ananthanarayan:2014sea,Ellis:2019zex,Ellis:2020ljj,Hernandez-Juarez:2021mhi,Hernandez-Juarez:2022kjx,Ellis:2022zdw,Ellis:2023zim}.}. In a previous study \cite{Jahedi:2022duc}, we explored the assessment of dim-8 nTGCs through $Z\gamma$ production at $e^+e^-$ colliders. In the publication at hand, we shift our focus to the examination of nTGCs via $ZZ$ production at the same type of colliders. This investigation runs in parallel with the aim of achieving a comprehensive understanding of future nTGCs estimations at the $e^+e^-$ colliders.

As our intention is to investigate the precision of nTGCs, $e^+ e^-$ colliders are in the focus of our interest because of the absence of QCD background. On the other hand, as electron and positron are fundamental particles, the non-existence of PDF uncertainties in the initial beams is useful to estimate an NP signal in this much cleaner background. Moreover,
the accessibility of partially polarized beams is also beneficial for suppressing the SM background and allowing the NP signal dominating over SM background. Henceforth, we will inspect our analysis on one of the proposed $e^+e^-$ colliders such as the Compact Linear Collider (CLIC) \cite{Aicheler:2018arh} with highest center of mass (CM) energy  along with 1000 $\rm fb^{-1}$ integrated luminosity considering maximum beam polarization combination. We will compare our results with the existing collider bound prevailed at the LHC \cite{ATLAS:2011nmx,ATLAS:2013way,CMS:2015wtk,CMS:2016cbq} and the precursor of this, the $e^+e^-$ collider LEP \cite{L3:2004hlr,OPAL:2003gfi}. The tightest bound for three dim-8 effective couplings so far has been given by ATLAS at $\sqrt{s}$ = 13 TeV and 36.1 $\rm fb^{-1}$ integrated luminosity \cite{ATLAS:2018nci} whereas for one dim-8 coupling, CMS has provided the most stringent bound at $\sqrt{s}$ = 13 TeV and 36.1 $\rm fb^{-1}$ integrated luminosity \cite{CMS:2020gtj}. In this analysis, we aim to provide better constrains on nTGCs compared to existing LHC limits.

In addition to the usual cut-based (binned) analysis, the optimal observable technique (OOT) \cite{Atwood:1991ka, Davier:1992nw, Diehl:1993br, Gunion:1996vv} is another statistical method to provide the sensitivity of NP couplings.  OOT has been extensively used to constrain top-quark couplings \cite{Grzadkowski:1996pc,Grzadkowski:1997cj,Grzadkowski:1998bh,Grzadkowski:1999kx,Grzadkowski:2000nx,Bhattacharya:2023mjr} and Higgs couplings \cite{Hagiwara:2000tk,Dutta:2008bh} in case of $e^+e^-$ colliders. The investigation of top quark interaction at $\gamma \gamma$ colliders \cite{Grzadkowski:2003tf,Grzadkowski:2004iw,Grzadkowski:2005ye}, the measurement of top-Yukawa couplings at the LHC \cite{Gunion:1998hm}, at the muon collider \cite{Hioki:2007jc} and $e \gamma$ colliders \cite{Cao:2006pu} have also been done in the context of OOT. Recent work on OOT includes the $Z$ couplings of heavy charged fermions at $e^+e^-$ colliders \cite{Bhattacharya:2021ltd} as well as exploring NP effects in flavor physics scenarios \cite{Bhattacharya:2015ida,Calcuttawala:2017usw,Calcuttawala:2018wgo}. In this work, we opt for this powerful technique to estimate the sensitivity of NP couplings by using the above described CM energy and integrated luminosity at CLIC.

\noindent
This paper is arranged as follows: In section~\ref{sec:dim8}, we discuss the SMEFT framework relevant for our study. A short overview of OOT is sketched in section~\ref{sec:oot}. In section~\ref{sec:coll}, we briefly discuss the collider analysis for a specific final state signal. We then show our elaborated numerical analyses in section~\ref{sec:result}. Finally, we conclude our discussion in section~\ref{sec:con}.
\section{ SMEFT framework for $ZZV$ couplings}
\label{sec:dim8}
The dim-8 effective lagrangian involving nTGCs in the framework of SMEFT can be written as \cite{Degrande:2013kka} 
\beq
\mathcal{L^{\tt nTGCs}} = \mathcal{L}_{\tt SM} + \sum_{i}\frac{C_i}{\Lambda^4}\left( \mathcal{O}_i +\mathcal{O}_i^{\dagger}  \right),
\eeq
where $\mathcal{O}_i$ are the dim-8 SMEFT operators that consist of SM fields and obey the SM gauge symmetry. $C_i$ are the dimensionless Wilson coefficients through which NP effects can be understood and $\Lambda$ is the scale of NP. Dim-8 operators that contribute to the $ZZV$ couplings are given by \cite{Degrande:2013kka}:
\bea \begin{aligned}
	\mathcal{O}_{\tilde B W}&=iH^{\dagger}\tilde{B}_{\mu\nu}W^{\mu\rho}\{D_{\rho},D^{\nu}\}H,\\
	\mathcal{O}_{ B W}&=iH^{\dagger}B_{\mu\nu}W^{\mu\rho}\{D_{\rho},D^{\nu}\}H,\\
	\mathcal{O}_{ W W}&=iH^{\dagger}W_{\mu\nu}W^{\mu\rho}\{D_{\rho},D^{\nu}\}H,\\
	\mathcal{O}_{ B B}&=iH^{\dagger}B_{\mu\nu}B^{\mu\rho}\{D_{\rho},D^{\nu}\}H,\\
	\label{eq:dim8}
\end{aligned}
\eea
where $D_{\mu}=(\partial_{\mu}-i g W_{\mu}^{i}\sigma^{i}-i \frac{g'}{2}B_{\mu}Y)$ is the conventional covariant derivative, $H$ is the SM Higgs doublet, and $B_{\mu\nu}$ and $W_{\mu\nu}$ are the $U(1)$ and $SU(2)$ gauge field strength tensors, respectively. The definitions are as follows:
\begin{align}
	B_{\mu\nu}&=\partial_{\mu}B_{\nu}-\partial_{\nu}B_{\mu},\\
W_{\mu \nu}&=\sigma^{i}\left(\partial_{\mu}W_{\nu}^{i}-\partial_{\nu}W_{\mu}^{i}+g \epsilon_{ijk}W_{\mu}^{j}W_{\nu}^{k}\right), \quad \rm{with \, \vevof{\sigma^{i}\sigma^{j}}}=\frac{\delta^{ij}}{2}.
\end{align}
The dual tensors are defined by $\tilde{X}_{\mu \nu}=\frac{1}{2}\epsilon_{\mu \nu \rho \sigma} X^{\rho \sigma}$. In Eq.~\eqref{eq:dim8}, $B_{\mu \nu}$ is CP odd and its dual tensor $\tilde{B}_{\mu \nu}$ is CP even due to $\epsilon_{\mu \nu \rho \sigma}$. When  CP operator acts on  $\mathcal{O}_{\tilde{B}W}$, we find that $iH^{\dagger}$ and $W_{\mu \nu}$ are CP odd, while $\tilde{B}_{\mu \nu}$, $D_{\mu}$, and $H$ are CP even. Therefore, $\mathcal{O}_{\tilde{B}W}$ is CP even, whereas the other three operators are CP odd as they do not contain $\epsilon_{\mu \nu \rho \sigma}$. These CP odd operators contribute to the electric dipole moment (EDM) of the electron. The upper bound for these dim-8 CP-violating effective couplings, as determined by the EDM constraint \cite{ACME:2018yjb}, is roughly of the order of $10^{-3}$ $\rm{TeV^{-4}}$. This upper bound is determined through $W$ mediated one-loop process originating from the $Q_{leW^2 H}$ operator. Subsequently, this limit is translated on the $\mathcal{O}_{WW}$ operator using Eq.~\eqref{eq:dim8.coup}. Therefore, we anticipate a similar order of bounds for the other three couplings. Notably, this constraint aligns with the collider constraints we elaborate in Table \ref{tab:95cl}. The operators outlined in Eq.~\eqref{eq:dim8} can be expressed as combination of the operators from the complete dim-8 SMEFT basis listed in \cite{Murphy:2020rsh}. In terms of couplings, the relations are as follows:
\begin{align}
\nonumber
C_{\tilde{B}W}=&-ig'C^{(2)}_{W^2BH^2}+\frac{1}{2}(C^{(6)}_{WBH^2D^2}-iC^{(5)}_{WBH^2D^2})-2\lambda C_{WBH^4}^{(2)}+2 C_{WBH^2D^2}^{(2)}\\\nonumber
&-2( y^q_u C^{(2)}_{quWBH}+ y^q_d C^{(2)}_{qdWBH}+ y^e C^{(2)}_{leWBH}),\\\nonumber
C_{BW}=&-igC^{(1)}_{W^2BH^2}+\frac{1}{2}(C^{(4)}_{WBH^2D^2}-iC^{(3)}_{WBH^2D^2})-2\lambda C_{WBH^4}^{(1)}+2C_{WBH^2D^2}^{(1)}\\
\label{eq:dim8.coup}
&-2( y^q_u C^{(1)}_{quWBH}+ y^q_d C^{(1)}_{qdWBH}+ y^e C^{(1)}_{leWBH}),\\\nonumber
C_{WW}=&-ig'C^{(1)}_{W^2BH^2}-igC^{(1)}_{W^3H^2} +2 C_{W^2 H^2 D^2}^{(1)}-2\lambda C_{W^2 H^4}^{(1)}+2C_{W^2H^2D^2}^{(2)}
\\\nonumber
&-2( y^q_u C^{(1)}_{quW^2H}+ y^q_d C^{(1)}_{qdW^2H}+ y^e C^{(1)}_{leW^2H}),\\\nonumber
C_{BB}=&2 (C_{B^2 H^2 D^2}^{(1)}-\lambda C^{(1)}_{B^2 H^4}+C_{B^2H^2D^2}^{(2)}- y^q_u C^{(2)}_{quB^2H}
- y^q_d C^{(2)}_{qdB^2H}- y^e C^{(2)}_{leB^2H}),\\\nonumber
\end{align}
where $\lambda$ is quartic Higgs coupling, and $ y^q_u$,  $y^q_d$ and  $y^e$ are the Yukawa couplings corresponding to up-type quarks, down-type quarks and charged leptons, respectively. The full decomposition of a single operator ($\mathcal{O}_{WW}$) has been done in Appendix \ref{sec:opsexpan} to establish the connection between the operators in Eq.~\eqref{eq:dim8} with the the operators in the universal basis mentioned in \cite{Murphy:2020rsh}. Other operators follow a similar decomposition.

There is no contribution from dim-6 SMEFT operators to the nTGCs at tree level but they can arise at one-loop level. However, if we compare the order of the contribution to the production cross-section between one loop dim-6 operators and tree level dim-8 operators, we can evaluate that the order of contribution is $\mathcal{O}(\frac{\alpha_{\tt EM} s}{4 \pi \Lambda^2})$ in case of dim-6 operators whereas for dim-8 operators the contribution would be $\mathcal{O}(\frac{s v^2}{\Lambda^4})$. Therefore, the contribution from tree level dim-8 nTGCs operators to the production cross-section  overpowers the contribution of one-loop dim-6 operators in the limit of $\Lambda < \sqrt{\frac{4 \pi}{\alpha_{\tt EM}}}v\sim 10$ TeV. 

Considering both dim-6 and dim-8 operator contributions, the effective lagrangian with anomalous couplings is given by \cite{Gounaris:1999kf,Degrande:2013kka}
\begin{align}
	\nonumber
	\mathcal{L}_{\tt EFT}=&\frac{g_e}{m_Z^2}\bigg[-\big\{f_{4}^{\gamma}(\partial_{\mu}F^{\mu \nu})+f_{4}^{Z}(\partial_{\mu}F^{\mu \beta})\big\}Z_{\alpha} (\partial^{\alpha} Z_{\beta})+\big\{f_{5}^{\gamma}(\partial^{\sigma}F_{\sigma \mu})+f_{5}^{Z}(\partial^{\sigma}Z_{\sigma \mu})\big\}\tilde{Z}^{\mu \beta} Z_{\beta}\\\nonumber
	&-\big\{h_{1}^{\gamma}(\partial^{\sigma}F_{\sigma \mu})+h_{1}^{Z}(\partial^{\sigma}Z_{\sigma \mu})\big\} Z_{\beta} F^{\mu \beta}-\big\{h_{3}^{\gamma}(\partial_{\sigma}F^{\sigma \rho})+h_{3}^{Z}(\partial_{\sigma}Z^{\sigma \rho})\big\} Z^{\alpha} \tilde{F}_{\rho \alpha}\\\nonumber
	&-\bigg\{\frac{h^{\gamma}_2}{m^2_Z}(\partial_{\alpha} \partial_{\beta} \partial^{\rho} F_{\rho \mu}) +\frac{h^{Z}_2}{m^2_Z} \big(\partial_{\alpha} \partial_{\beta}(\Box + m_Z^2)Z_{\mu}\big)\bigg\}Z^{\alpha} F^{\mu \beta}-\bigg\{\frac{h^{\gamma}_4}{2m^2_Z}(\Box \partial^{\alpha} F^{\rho \alpha})\\
	& +\frac{h^{Z}_4}{2m^2_Z} \big((\Box + m_Z^2) \partial^{\sigma} Z^{\rho \alpha}\big)\bigg\}Z_{\alpha} \tilde{F}_{\mu \beta}\bigg],
\end{align}
where $\tilde{Z}_{\mu \nu}=\frac{1}{2}\epsilon_{\mu \nu \rho \sigma} Z^{\rho \sigma}$ and $Z_{\mu \nu}=\partial_{\mu} Z_{\nu}-\partial_{\nu} Z_{\mu}$. Here, $f_3^{V}, \, f_4^{V}, $ and $ f_5^{V}$ ($V=\gamma, Z$) are the anomalous dim-6 couplings whereas $h_3^{V}, \, h_4^{V}, $ and $h_5^{V}$ are the anomalous dim-8 couplings. In terms CP conserving dim-8 nTGC, the anomalous couplings are written as \cite{Degrande:2013kka}:

\begin{align}
	h_3^Z&=\frac{v^2 m_Z^2 C_{\tilde{B}W}}{4 c_w s_w \Lambda^4},\\
	h_4^Z&=h_3^{\gamma}=h_4^{\gamma}=0,\\\nonumber
\end{align}

whereas, in case of CP violating scenario, the couplings has the following forms \cite{Degrande:2013kka}:

\begin{align}
	h_1^Z&=\frac{m_Z^2 v^2\left(-c_w s_w C_{WW}+C_{BW}(c_w^2-s_w^2)+4 c_w s_w C_{BB}\right)}{4 c_w s_w \Lambda^4}, \\
	h_2^Z&=h_2^{\gamma}=0, \\
	h_1^{\gamma}&=\frac{m_Z^2 v^2\left(s_w^2 C_{WW}-2 c_w s_wC_{BW}+4 c_w^2 C_{BB}\right)}{4 c_w s_w \Lambda^4}.
\end{align}

Experimental searches for these nTGCs have been going on in the different colliders but no evidence has been found so far. As discussed above, searches for nTGCs have been carried out at LEP  and Tevatron but most stringent bounds come from the ATLAS and CMS experiments at LHC. The ATLAS experiment has put most stringent bounds  on $C_{ \tilde{B} W}/\Lambda^4$, $C_{ B W}/\Lambda^4$ and $C_{BB}/\Lambda^4$ couplings through the $pp \to Z \gamma \to \nu \bar{\nu} \gamma$ channel at CM energy $\sqrt{s}$ = 13 TeV with integrated luminosity of 36.1 $\rm fb^{-1}$ at the LHC \cite{ATLAS:2018nci} whereas for the $C_{ W W}/\Lambda^4$ coupling, the tightest limit is given by the CMS experiment through the $pp \to Z Z \to 4  \ell$ channel at $\sqrt{s}$ = 13 TeV with integrated luminosity 137 $\rm fb^{-1}$ \cite{CMS:2020gtj}. The expected 95\% C.L. on dim-8 nTGCs from ATLAS and CMS experiments are presented in Table \ref{tab:limit}.

\begin{table}[htb!]
	\centering
	{\renewcommand{\arraystretch}{1.3}%
		\begin{tabular}{ |c|c|c| } 
			\hline
			\multicolumn{1}{|c}{Couplings} &
			\multicolumn{2}{|c|}{ 95\% C.L.}\\
			\cline{2-3}
			$(\rm TeV^{-4})$ & $pp \to Z\gamma \to \nu \bar{\nu} \gamma$ (ATLAS) & $pp \to ZZ \to 4 \ell$ (CMS) \\[0.6ex]
			\hline\hline
			$C_{\tilde B W}/\Lambda^4$ & $(-1.10, \, +1.10)$ & $(-2.30, \, +2.50)$ \\ [0.6ex] 
			$C_{W W}/\Lambda^4$ & $(-2.30, \, +2.30)$ & $(-1.40, \, +1.20)$ \\ [0.6ex]
			$C_{ B W}/\Lambda^4$ & $(-0.65, \, +0.64)$ &$(-1.40, \, +1.30)$  \\ [0.6ex]
			$C_{B B}/\Lambda^4$ & $(-0.24, \, +0.24)$ & $(-1.20, \, +1.20)$ \\  [0.6ex]
			\hline
	\end{tabular}}
	\caption{Statistical limits (95\% C.L.) from ATLAS \cite{ATLAS:2018nci} and CMS \cite{CMS:2020gtj} experiments on different dim-8 nTGCs at LHC.}
	\label{tab:limit}
\end{table}

In the following analysis, we estimate the statistical limits of dim-8 nTGCs considering the highest CM energy 3 TeV and the maximum polarization combination $\{P_{e^-}:P_{e^+}=\pm80\%:0\%\}$ of the incoming beams with $\mathfrak{L}_{\tt int}$ = 1000 $\rm fb^{-1}$ integrated luminosity for CLIC.

\section{Optimal Observable Technique}
\label{sec:oot}
The optimal observable technique (OOT) is a credible tool to determine the statistical limit of any NP coupling in a prudent way. Here, we concisely sketch the mathematical framework of OOT which has already been elaborated in detail \cite{Diehl:1993br,Gunion:1996vv}. In general, any observable ({\it e.g.} differential cross-section) that gets contribution from the SM and BSM can be presented as
\beq
\mathcal{O}(\phi)=\frac{d\sigma_{\tt theo}}{d\phi} = \sum_i g_i f_i(\phi) \,,
\label{eq:expnd1}
\eeq
where $ \phi $ is a phase-space variable, the coefficients $g_i$ are the functions of NP couplings and several numerical constants, and  $ f_i $ are the  linearly-independent functions of the phase space variable $\phi$. In this analysis, as we will be probing the 2 $\rightarrow$ 2 scattering process ($e^+e^-\rightarrow ZZ$), the cosine of the CM frame ($\cos\theta$) is the phase-space variable $\phi$ in our consideration. In principle, $\phi$ can be chosen to be any other variable as well, depending on the process of interest.

The determination of $g_i$ can be estimated by using a suitable weighting function ($w_i(\phi)$):

\beq
g_i=\int w_i(\phi) \mathcal{O}(\phi) d\phi.
\eeq

In principle, different choices of $w_i(\phi)$ are possible, but there is a distinctive choice for which the covariance matrix ($V_{ij}$) is optimal in a sense that the statistical uncertainties in NP couplings are minimized. For this choice, $V_{ij}$ follows:

\beq
V_{ij} \propto \int w_i({\phi})w_j({\phi}) \mathcal{O}({\phi}) d\phi.
\eeq

Therefore, the weighting functions considering the optimal condition $\delta V_{ij}=0$  are 

\beq
w_i(\phi)=\frac{M_{ij}^{-1}f_j(\phi)}{\mathcal{O}(\phi)},
\eeq

where,

\beq
M_{ij} =\int \frac{f_i(\phi) f_j(\phi)}{\mathcal{O}(\phi)}d\phi.
\label{eq:mij}
\eeq

Accordingly, the optimal covariance matrix takes form 
\beq
V_{ij} =\frac{ M_{ij}^{-1} \sigma_T}{N}= \frac{M^{-1}_{ij}}{\mathfrak{L}_{\tt int}}\, ,
\label{eq:covmat1}
\eeq
where $\sigma_T=\int \mathcal{O}(\phi) d\phi$ and $N=\sigma_T \mathfrak{L}_{\tt int}$ is total number of events. $\mathfrak{L}_{\tt int}$ is the integrated luminosity.

The $\chi^2$ function that determines the optimal limit of NP couplings is defined as 

\beq
\chi^2= \sum_{\{i,j\}=1}^{n} (g_i -g_i^0) (g_j -g_j^0) \,  \left( V^{-1}\right)_{ij},
\label{eq:chi2}
\eeq
where the $g^0$ are the {\it `seed values'} that depend on the specific NP model. The limit provided by $\chi^2 \le n^2$ corresponding to $n\sigma$ standard deviations from a seed values ($g^0$)  is the optimal limit for any NP couplings by admitting that the covariance matrix ($V_{ij}$) is minimal. Using the definition of the $\chi^2$ function in Eq.~\eqref{eq:chi2}, the optimal limit on the NP couplings have been explored in the following sections.

\section{Collider analysis}
\label{sec:coll}
\begin{figure}[htb!]
	$$
	\includegraphics[height=5.2cm, width=7.2cm]{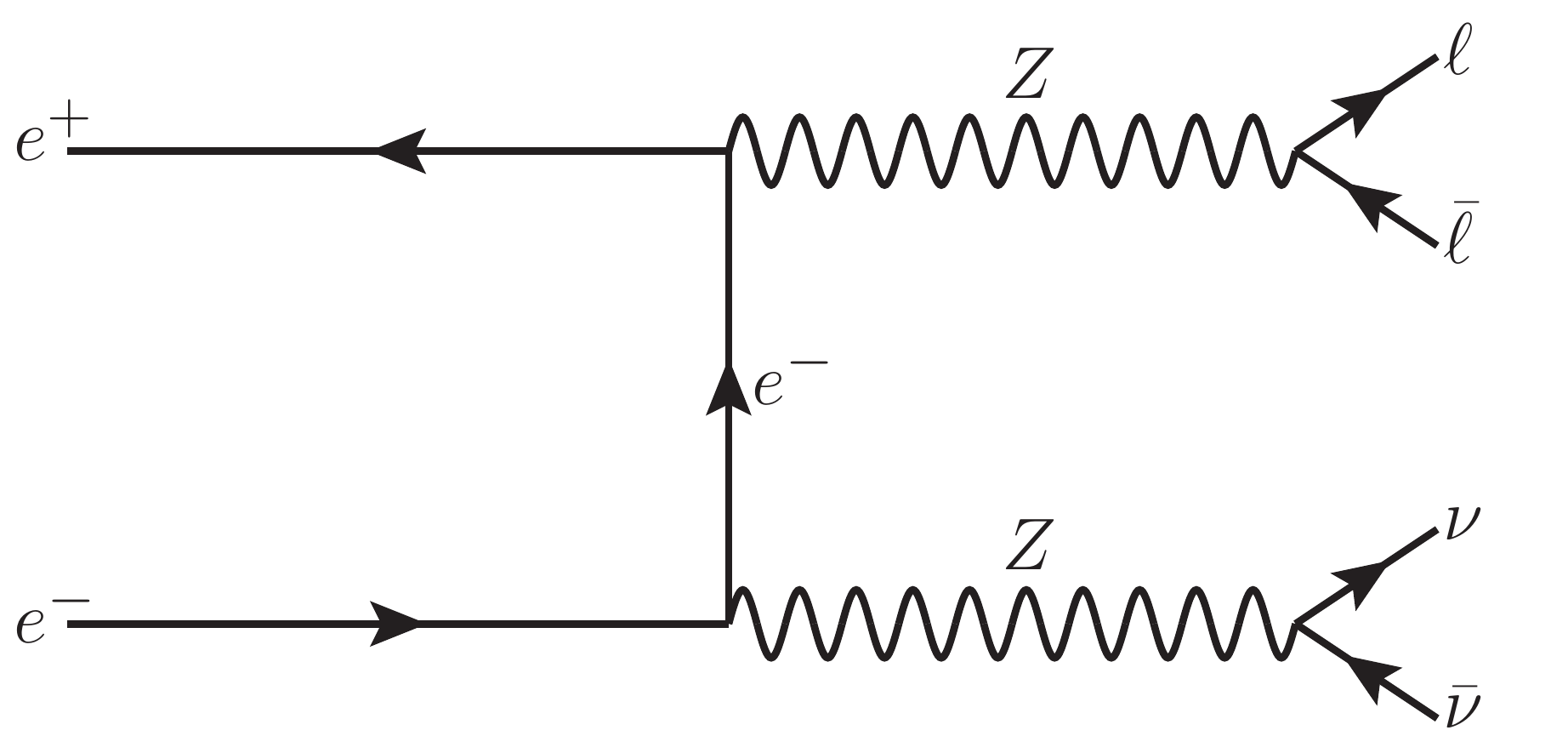}
	\includegraphics[height=5.2cm, width=7.2cm]{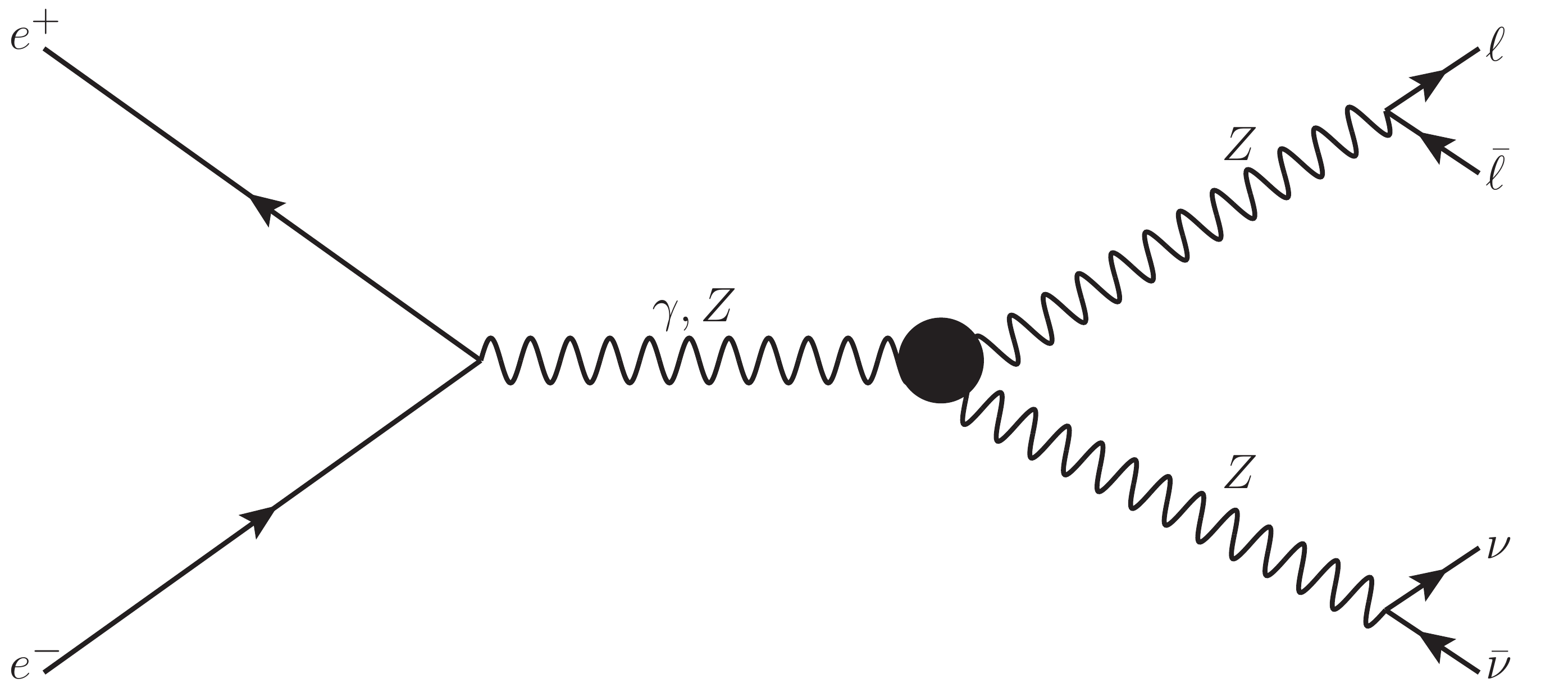}
	$$
	\caption{$Z$ boson pair production and subsequent decay to {$\tt OSL$} + missing energy ($E_{\tt miss}$); left: SM, right: dim-8 SMEFT.} 	
	\label{fig:feyn.diag}
\end{figure}

In this section, we will briefly discuss the strategy of  estimating signal-background events for $ZZ$ production by reducing the non-interfering SM backgrounds for a specific final state signal of our consideration. The production and subsequent decay final state signal (see figure \ref{fig:feyn.diag}) for the analysis is given by
\beq
e^+ e^- \to Z Z, \, Z \to \ell^+ \ell^- \,(\ell=e,\mu), \, Z\to \nu \bar{\nu}.
\eeq 
This final state signal provides missing energy through the SM neutrinos which is advantageous to discriminate other non-interfering SM backgrounds from the $ZZ$ production. On top of that, as the Z decay to neutrinos has a higher branching ratio compared to charged leptons, this final state produces more statistics. The final state signal process of our interest is opposite sign lepton ($\tt OSL$) + missing energy ($E_{\tt miss}$). Dominant non-interfering SM backgrounds arise from 2-body $WW$ and $\ell^+\ell^-(\ell=\tau, \mu)$. There are 3-body non-interfering backgrounds like $WWZ$, $\nu \nu Z$ and $\ell \ell Z$ but their contribution is $\le 0.1\%$ compared to $WW$. Using a suitable cut on $\Delta R_{\ell \ell}$, the $\ell^+ \ell^-$ background can be fully reduced without harming the signal.  Therefore, in our following analysis, we try to reduce $WW$ as much as possible. In order to mimic the actual collider environment, the following selection criteria on leptons ($\ell=e, \mu$) and jets are implemented:
\begin{figure}[htb!]
	$$
	\includegraphics[height=6cm, width=7.6cm]{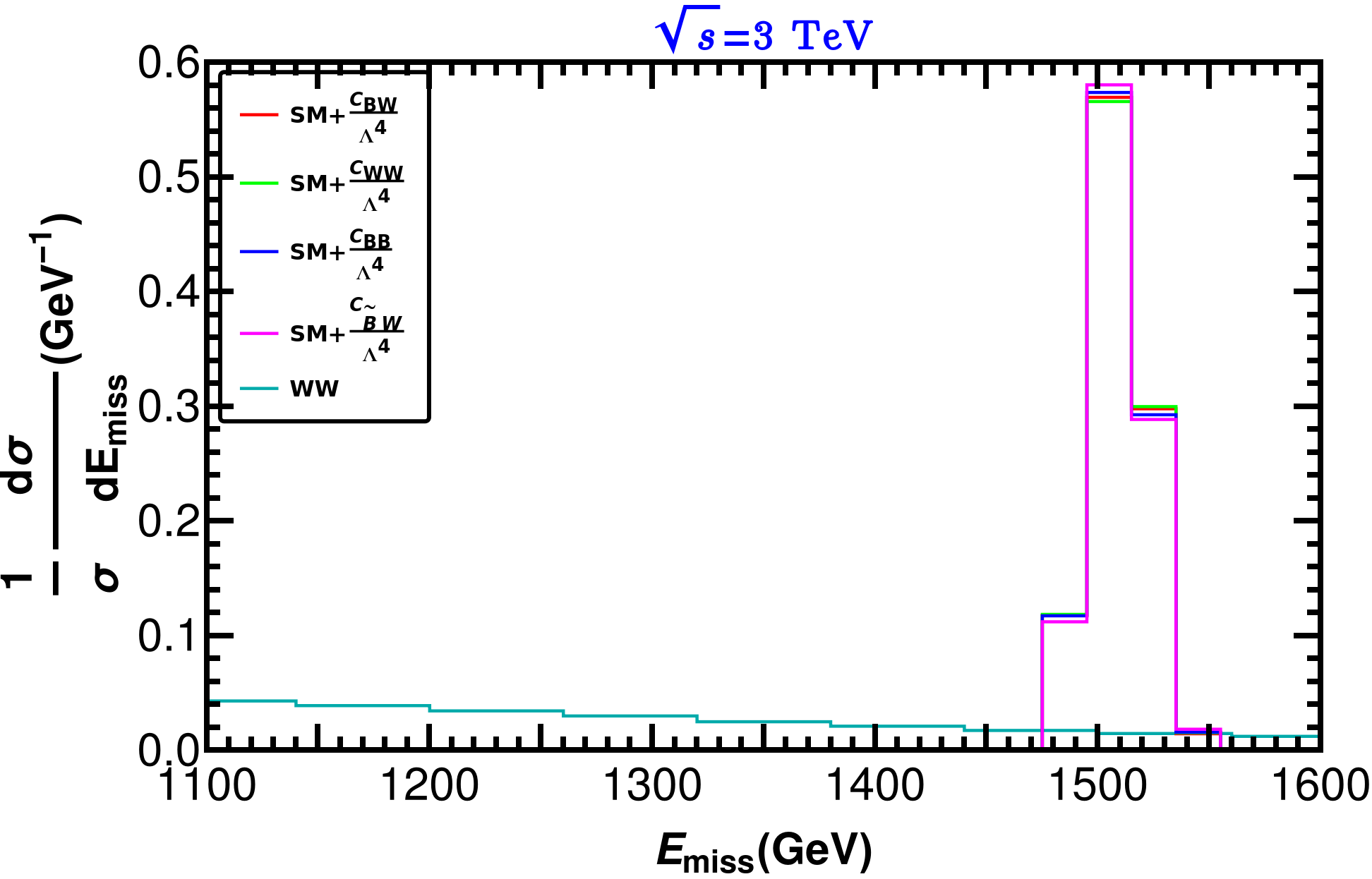}
	\includegraphics[height=6cm, width=7.6cm]{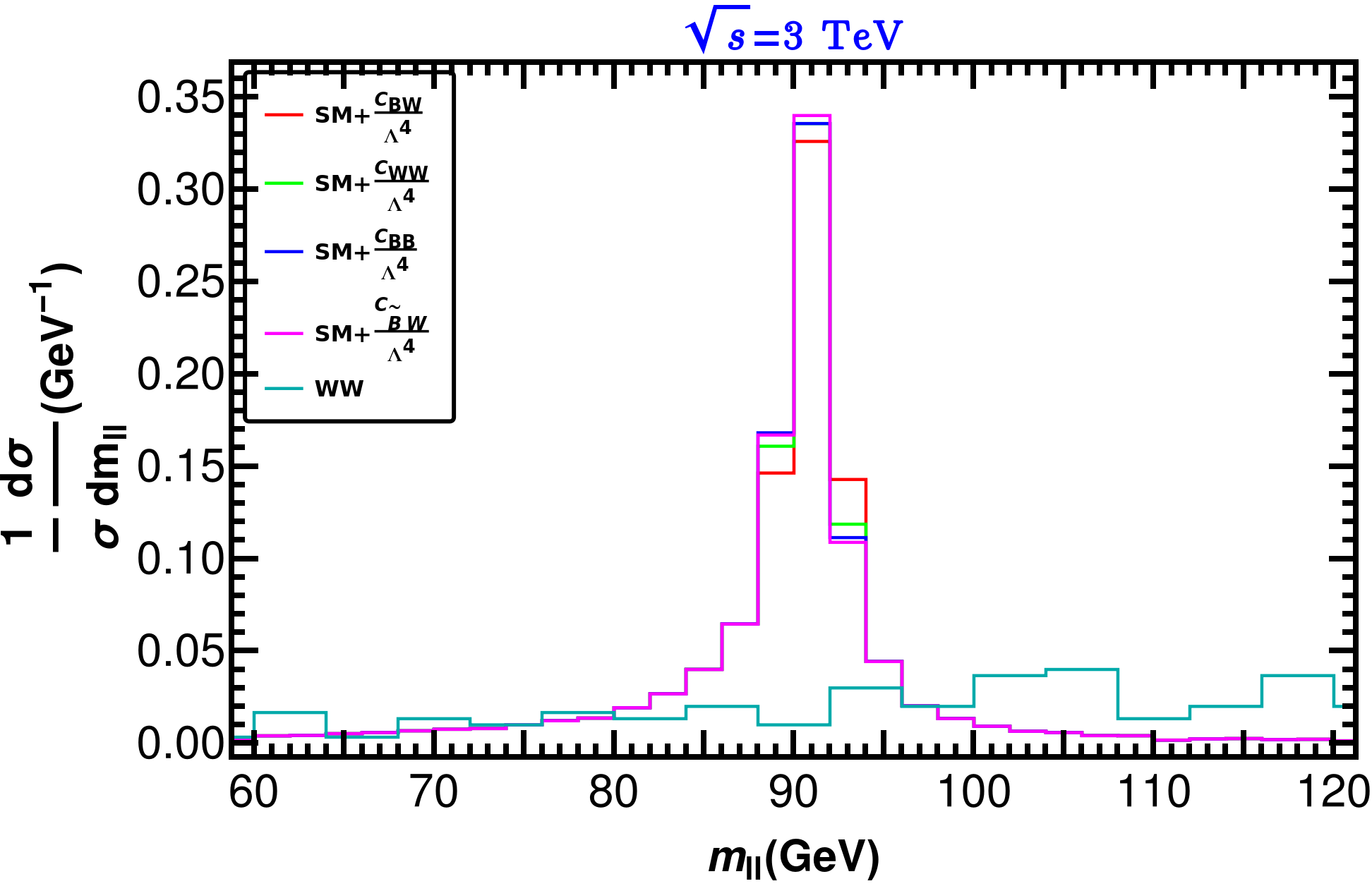}
	$$
	\caption{Normalized event distribution of $\tt OSL$ + missing energy for signal and non-interfering SM background with $\sqrt{s}= 3$ TeV, $C_{ij}=1$, $\Lambda=3.2$ TeV and unpolarized beam. Left: missing energy ($E_{\tt miss}$), right: invariant dilepton mass ($m_{\ell \ell}$). } 	
	\label{fig:evnt.dist}
\end{figure}
\begin{itemize}
	\item {\it Lepton} identification: Leptons required to have transverse momentum of $p^{\ell}_T > 10$ GeV and pseudorapidity $|\eta| < 2.5$. One lepton can be isolated from other leptons by the condition $\Delta R_{\ell \ell} \ge 0.2$. After that, leptons are separated from the jets by applying $\Delta R_{\ell j} > 0.4$ where $\Delta R=\sqrt{(\Delta \eta)^2+(\Delta \phi)^2}$ is the angular separation defined in azimuthal-pseudorapidity plane.  
	\item {\it Jet} identification: Jets are said to be isolated by imposing a transverse momentum cut of $p^{j}_T \ge 20$ GeV.   
\end{itemize}  
The kinematical variables are defined as follows:
\begin{itemize}
	\item {\it Missing energy ($E_{\tt miss}$)}: The energy carried by the missing particle in the final state can be written with the knowledge of CM energy $\sqrt{s}$ as
	\beq
	E_{\tt miss}=\sqrt{s}-\sum_{\ell,j, \gamma}E_i,
	\eeq
	where $E_i$ is the energy carried away by visible particles.
	\item {\it Invariant di-lepton mass ($m_{\ell \ell}$)}: The construction of an invariant di-lepton mass for two opposite sign leptons can be done by defining:
	\beq
	m_{\ell \ell}^2=(p_{\ell^+}+p_{\ell^-})^2,
	\eeq
	where $p_{\ell}$ is the 4-momentum of $\ell$.   
\end{itemize}

We generate signal and background events in $\tt Madgraph$ \cite{Alwall:2014hca}, then showered and analyzed through {\tt Pythia} \cite{Sjostrand:2014zea} and the detector simulation is done by ${\tt Delphes}$ \cite{deFavereau:2013fsa}. The $\tt UFO$ file that is feeded to {\tt Madgraph} is generated through $\tt FeynRules$ \cite{Christensen:2008py}. 

The normalized event distributions for $E_{\tt miss}$ and $m_{\ell \ell}$ are shown in figure~\ref{fig:evnt.dist}. The signal distributions shows interference between dim-8 couplings and the SM. If we look at signal event distributions for missing energy in case  of all nTGCs, the distribution peaks near $\sqrt{s}/2$ which is evident as one $Z$ decays to $\nu \bar{\nu}$. Therefore, a cut in the vicinity of $\sqrt{s}/2$ for $E_{\tt miss}$ reduces the irreducible non-interfering SM background significantly. Additionally, a sensible cut on $m_{\ell \ell}$ also helps us to reduce the other backgrounds even further. We observe that, with $1475 \, {\rm GeV}<E_{\tt miss}<1550$ GeV and $80 \, {\rm GeV}<m_{\ell \ell}< 98 \, {\rm GeV}$, all the signal distributions dominate over non-interfering SM backgrounds. This region of phase space is sensitive to study the nTGCs. Here, we note that the choice of the NP scale $\Lambda$ has been done in such a way that the EFT assumption remains valid. In the following discussion to estimate the sensitivity of dim-8 nTGCs, we will work in the region of phase space guided by the aforementioned kinematic cut values. 
\section{Results}
\label{sec:result}

\subsection{$ZZ$ production at the $e^+ e^-$ colliders}
\label{sec:cross}
Pair production of $Z$-boson at $e^+e^-$ colliders is primarily governed by the electron-mediated $t$-channel diagram within SM. The BSM contribution from dim-8 operators to the $ZZ$ production takes place via $s$-channel $\gamma$ and $Z$ mediation. Here, we consider that the NP contributions to the process $ e^+e^- \to ZZ$ other than nTGCs are negligible. In principle, tree level dim-6 SMEFT operators can contribute to the $ZZ$ production via $\gamma f \bar{f}$ and $Z f \bar{f}$ couplings. From the experimental measurements at LEP2 and LHC, the  $\gamma f \bar{f}$ ($\propto e_0$) coupling shows $\ll 0.1\%$ deviation from SM whereas for $Zf \bar{f}$ coupling ($\propto g$), the deviation is $\le0.1\%$ \cite{ParticleDataGroup:2022pth}\footnote{$e_0$ and $g$ are $U(1)_{em}$ and $SU(2)_L$ coupling constants, respectively.}. Therefore, we have neglected the contributions from dim-6 operators to $ZZ$ production.

The generic expression of the differential cross-section with partially polarized initial beams $\left(-1\le P_{e^{\pm}}\le+1\right)$ can be written as
\begin{align}
	\nonumber
	\frac{d\sigma(P_{e^+},\,P_{e^-})}{d\Omega} =& \frac{(1-P_{e^-})(1-P_{e^+})}4 \left( \frac{d\sigma}{d\Omega}\right)_{LL} +  \frac{(1-P_{e^-})(1+P_{e^+})}4 \left( \frac{d\sigma}{d\Omega}\right)_{LR} \\
	&\frac{(1+P_{e^-})(1-P_{e^+})}4 \left( \frac{d\sigma}{d\Omega}\right)_{RL} + \frac{(1+P_{e^-})(1+P_{e^+})}4 \left( \frac{d\sigma}{d\Omega}\right)_{RR},
	\label{eq:difcs}
\end{align}	
where $\left( \frac{d\sigma}{d\Omega}\right)_{ij}$ are the differential cross-sections with the electron and positron beams have helicities $i \in \{L,R\}$ and $j \in \{L,R\}$. $L$ and $R$ indicate left-handed and right-handed helicity of the initial beams, respectively\footnote{In general, left and right-handed helicity are denoted by `$-$' and `$+$' sign, respectively.}.
\begin{figure}[htb!]
	\begin{align*}
		\includegraphics[height=5cm, width=4.8cm]{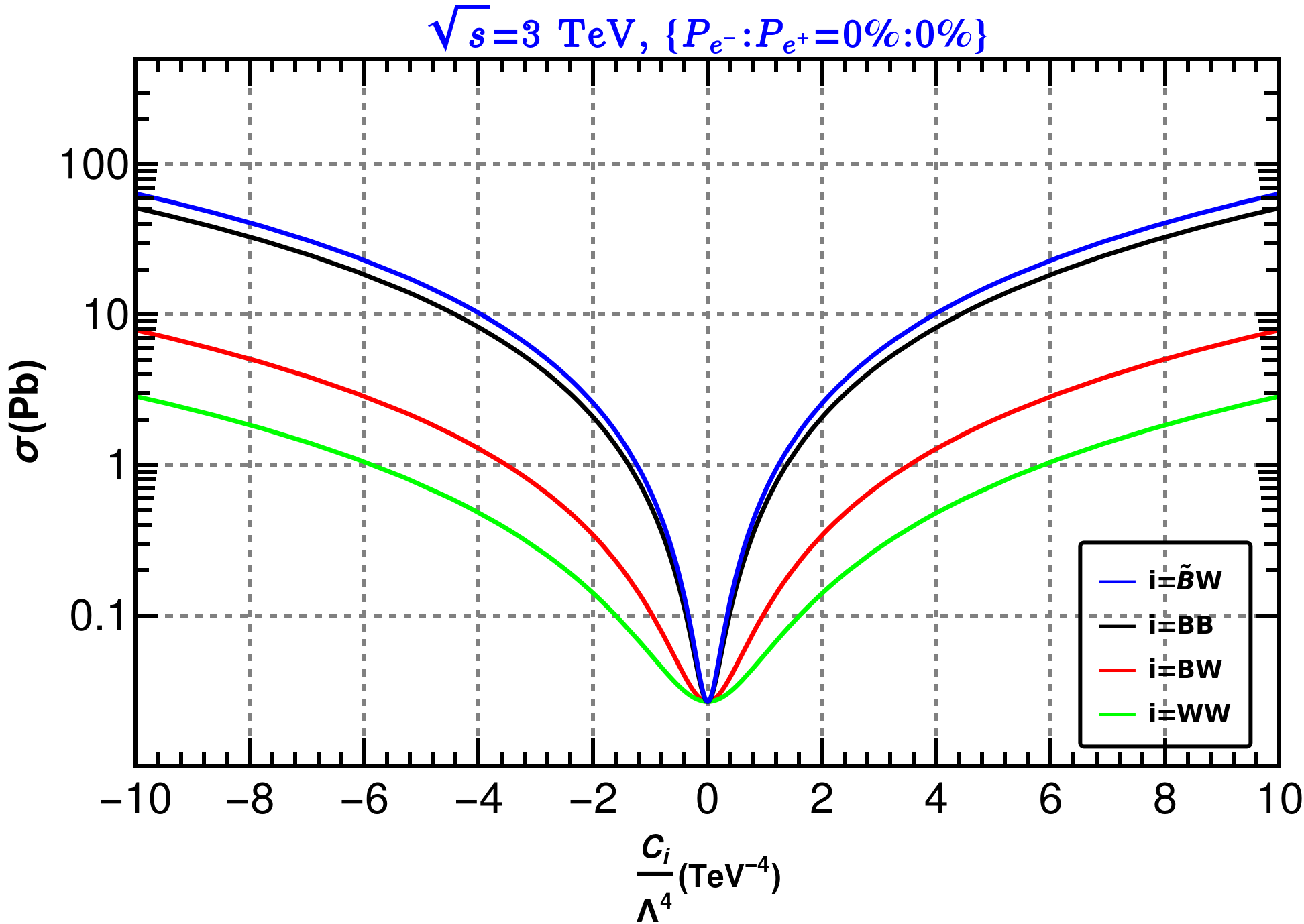} \quad
		\includegraphics[height=5cm, width=4.8cm]{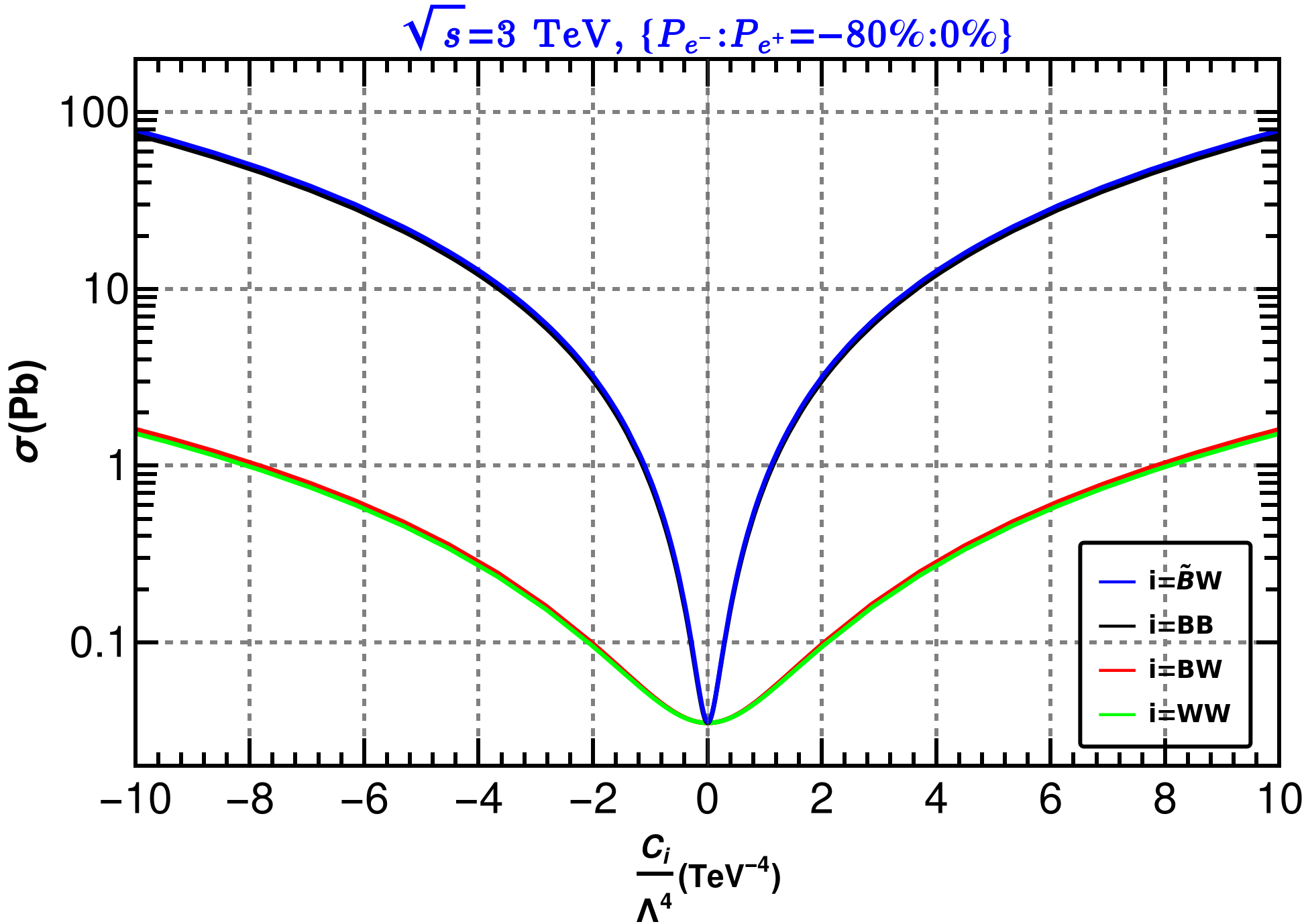} \quad		\includegraphics[height=5cm, width=4.8cm]{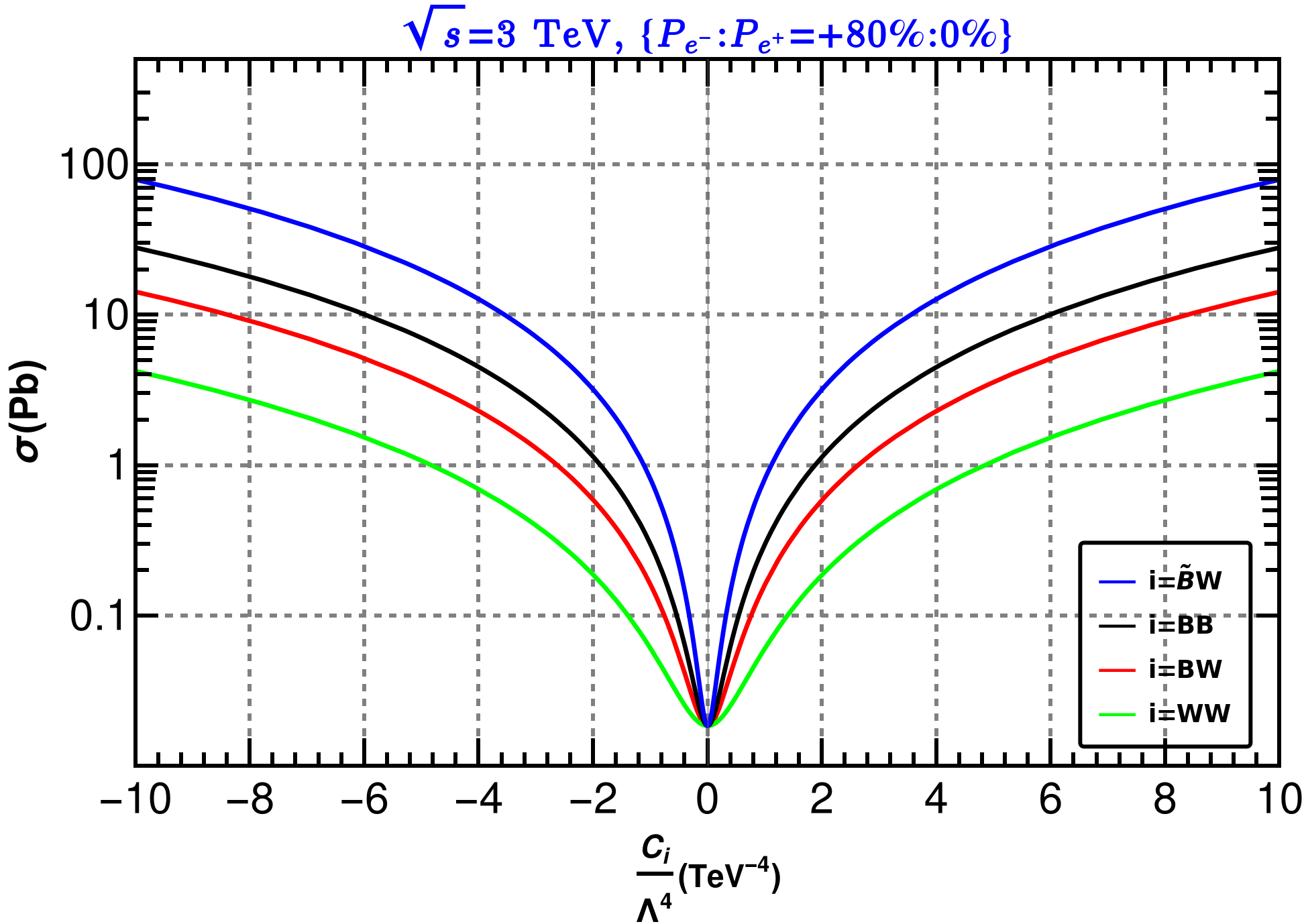} \quad
	\end{align*}
	\caption{ Variation of $ZZ$ cross-section with the different dim-8 effective couplings at the $e^+ \, e^-$ colliders with $\sqrt s$ = 3 TeV.  Left: $\{P_{e^-}:P_{e^+}=0\%:0\%\}$; Middle: $\{P_{e^-}:P_{e^+}=-80\%:0\%\}$; Right:$\{P_{e^-}:P_{e^+}=+80\%:0\%\}$.} 	
	\label{fig:xsec}
\end{figure}
The variation of the total cross-section with the dim-8 nTGCs for unpolarized and two different polarization combinations is shown in figure~\ref{fig:xsec}. The total amplitude of the process $e^+e^- \to ZZ$ is
\beq
|\mathcal{M}_{\tt tot}|^2=|\mathcal{M}_{\tt SM}|^2 +2 \mathcal{R} \left( \mathcal{M}_{\tt SM} \mathcal{M^*}_{\tt dim-8} \right) + |\mathcal{M}_{\tt dim-8}|^2.
\eeq
The dominant CP-even BSM contribution to the total cross-section arises from the interference term  $2 \mathcal{R} \left( \mathcal{M}_{\tt SM} \mathcal{M^*}_{\tt dim-8} \right)$ when $C_i/\Lambda^4$ are small. As the interference term is proportional to $C_i/\Lambda^4$, the cross-section shows asymmetric variation with $C_i/\Lambda^4$ in that small interval. In case of large $C_i/\Lambda^4$, the dominant contribution to the total cross-section comes from a pure BSM term i.e. $|\mathcal{M}_{\tt dim-8}|^2$ that is proportional to $\left(C_i/\Lambda^4\right)^2$. Therefore, in this range, behavior of the total cross-section is symmetric with $C_i/\Lambda^4$. CP violating dim-8 nTGCs do not interfere with the SM.  Therefore, their contributions may stand on a comparable footing or be suppressed by dim-10 and dim-12 SMEFT operators. However, in our analysis we assume that no NP effect except dim-8 operators contribute to $ZZ$ production. The beam polarization plays an important role to enhance or reduce the total cross-section. For $C_i/\Lambda^4 \, ( i=\tilde{B}W, BW, WW)$ couplings, the $ZZ$ cross-section is increased with $\{P_{e^-}:P_{e^+}=+80\%:0\%\}$ whereas for $C_{BB}/\Lambda^4$ coupling, the opposite polarization provides the enhancement. 
\subsection{Sensitivity of NP couplings using OOT}
\label{sec:sensitivity}
\begin{figure}[htb!]
	$$
	\includegraphics[height=5.5cm, width=7.2cm]{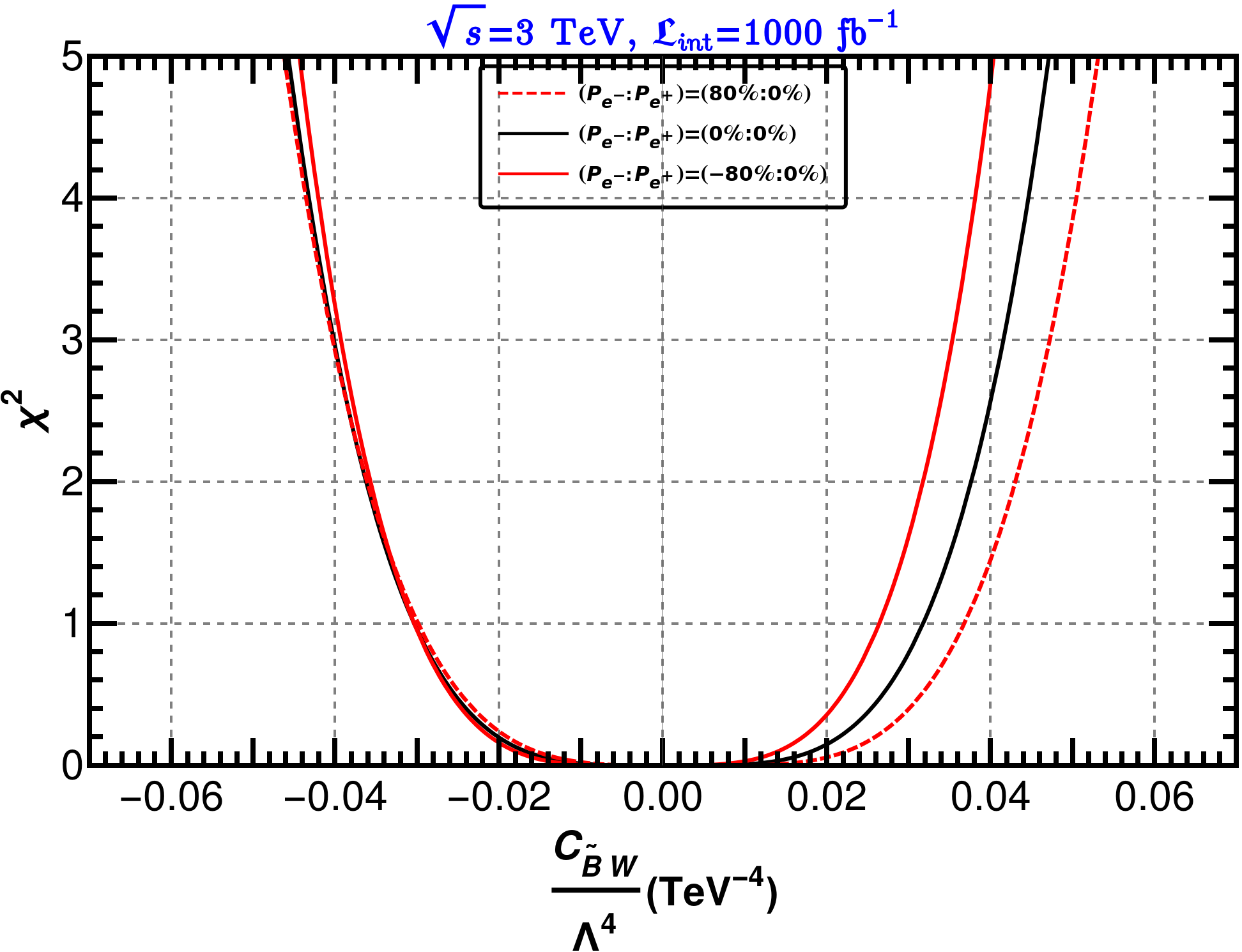}
	\includegraphics[height=5.5cm, width=7.2cm]{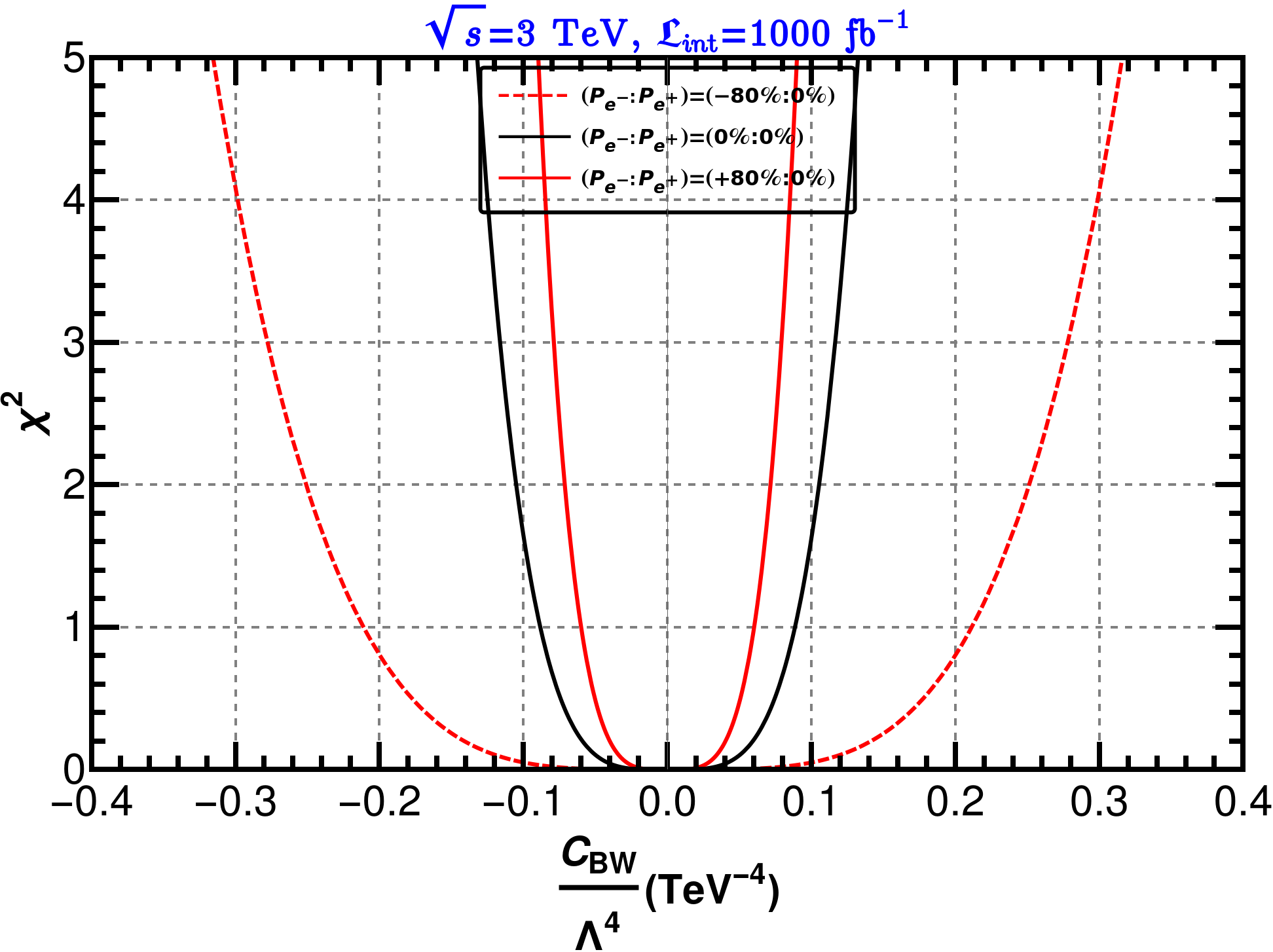}
	$$
	$$
	\includegraphics[height=5.5cm, width=7.2cm]{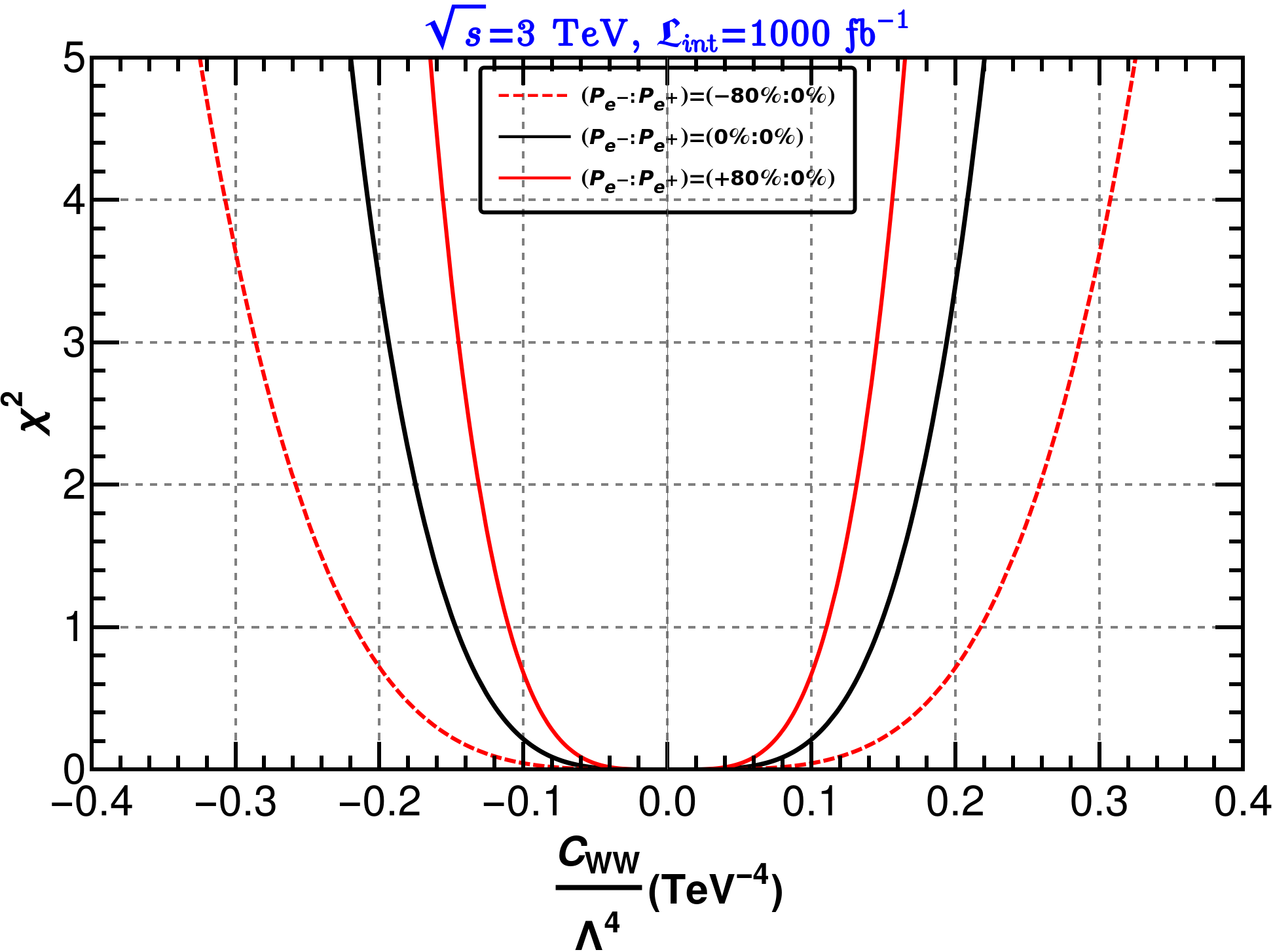}
	\includegraphics[height=5.5cm, width=7.2cm]{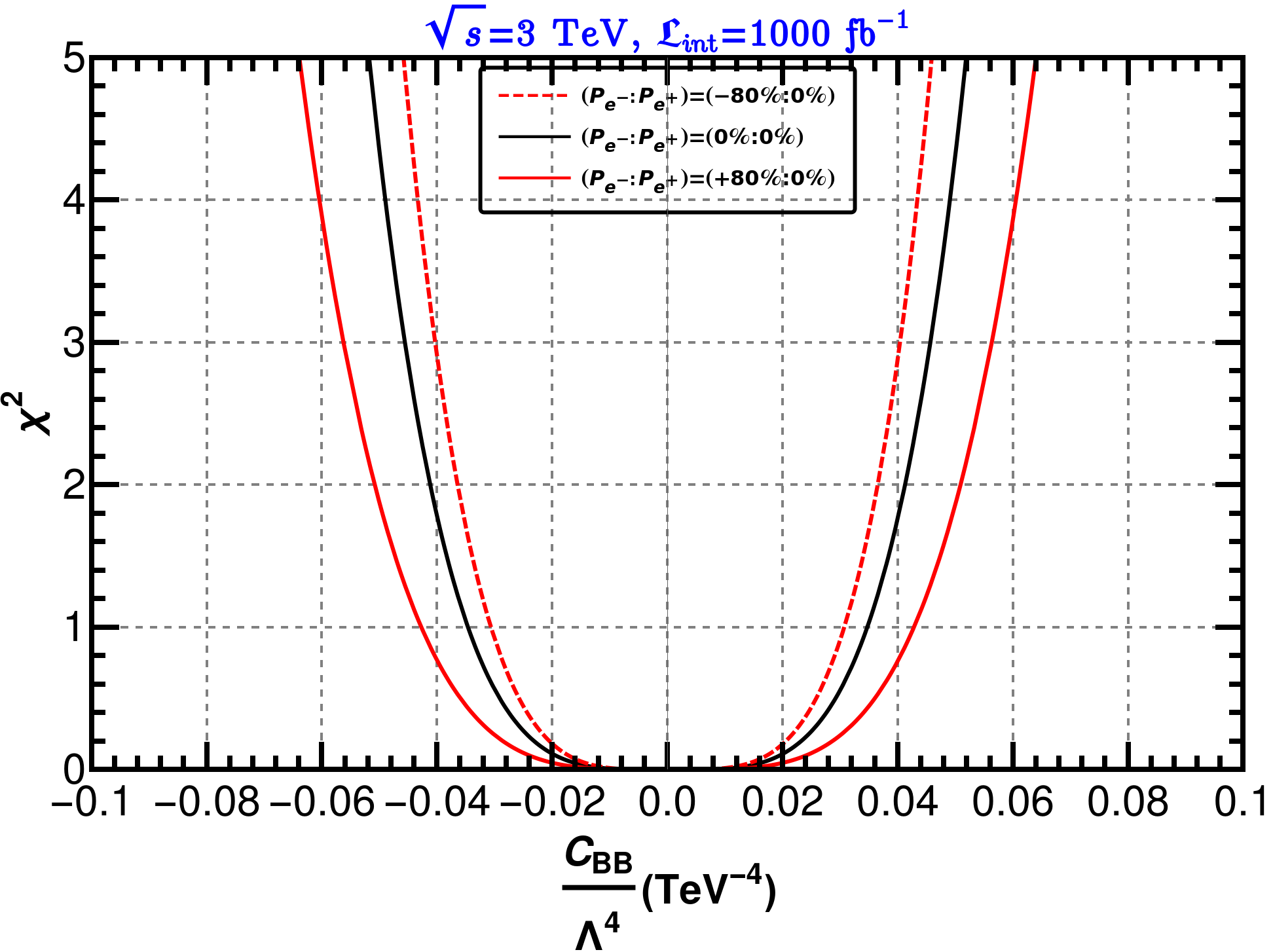}
	$$
	\caption{$\chi^2$ function of different dim-8 nTGCs for different choice of polarization combinations at CLIC. Polarization information is written in the inset. Top left: $C_{\tilde B W}/\Lambda^4$; top right: $C_{ B W}/\Lambda^4$; bottom left: $C_{W W}/\Lambda^4$; bottom right: $C_{BB}/\Lambda^4$. The CM energy is $\sqrt{s}=3$ TeV and integrated luminosity $\mathfrak{L}_{\tt int}=1000$ $\rm fb^{-1}$.} 	
	\label{fig:95cl1}
\end{figure}

The optimal sensitivity of NP couplings can be determined by using the $\chi^2$ function defined in Eq.~\eqref{eq:chi2}. We choose the CM energy $\sqrt{s}$ = 3 TeV and $\mathfrak{L}_{int}$ = 1000 $\rm{fb^{-1}}$ and then show $\chi^2$ variation with NP couplings for different choices of polarization combinations in figure~\ref{fig:95cl1}. The resulting statistical limits (95\%  C.L.) for each NP coupling with different beam polarization combinations   
have been tabulated in Table~\ref{tab:95cl}. Judicious choices of polarization combination help to estimate the bound on NP couplings in a most stringent way. $\{P_{e^-}:P_{e^+}=-80\%:0\%\}$ polarization combination provides the most optimal limit for $C_i/\Lambda^4 \, ( i=\tilde{B}W, BW, WW)$ couplings while for $C_{BB}/\Lambda^4$, the opposite polarization combination produces the best result. We can see that given the CM energy and integrated luminosity, CLIC provides much better constraints on NP coupings compared to current experimental bounds. For a careful choice of beam polarization, the most optimal limits of $C_{\tilde{B}W}/\Lambda^4$, $C_{BB}/\Lambda^4$, $C_{BW}/\Lambda^4$ and $C_{WW}/\Lambda^4$ couplings  are 29 (60), 6 (29), 8 (16) and 15 (9) times better than the latest experimental limits at ATLAS (CMS), respectively. The statistical sensitivity of NP couplings depends on their impact on the $ZZ$ production. For unpolarized beams, $C_{\tilde{B}W}/\Lambda^4$ maximally contributes, while $C_{WW}/\Lambda^4$ has the least influence. Notably, the contribution of $C_{\tilde{B}W}/\Lambda^4$ to $ZZ$ production surpasses that of $C_{WW}/\Lambda^4$ by a significant factor of thirty. This leads to $C_{\tilde{B}W}/\Lambda^4$ being constrainted approximately 7 times more tightly than $C_{WW}/\Lambda^4$ as evident from Table~\ref{tab:95cl}.

Using Eq.~\eqref{eq:dim8.coup}, excluding operators related to the fermionic current, the magnitude of the future sensitivity of bosonic operators related to dim-8 couplings are tabulated in Table~\ref{tab:95cl2} using above mentioned collider parameters.

\begin{figure}[htb!]
	$$
	\includegraphics[height=5.5cm, width=7.2cm]{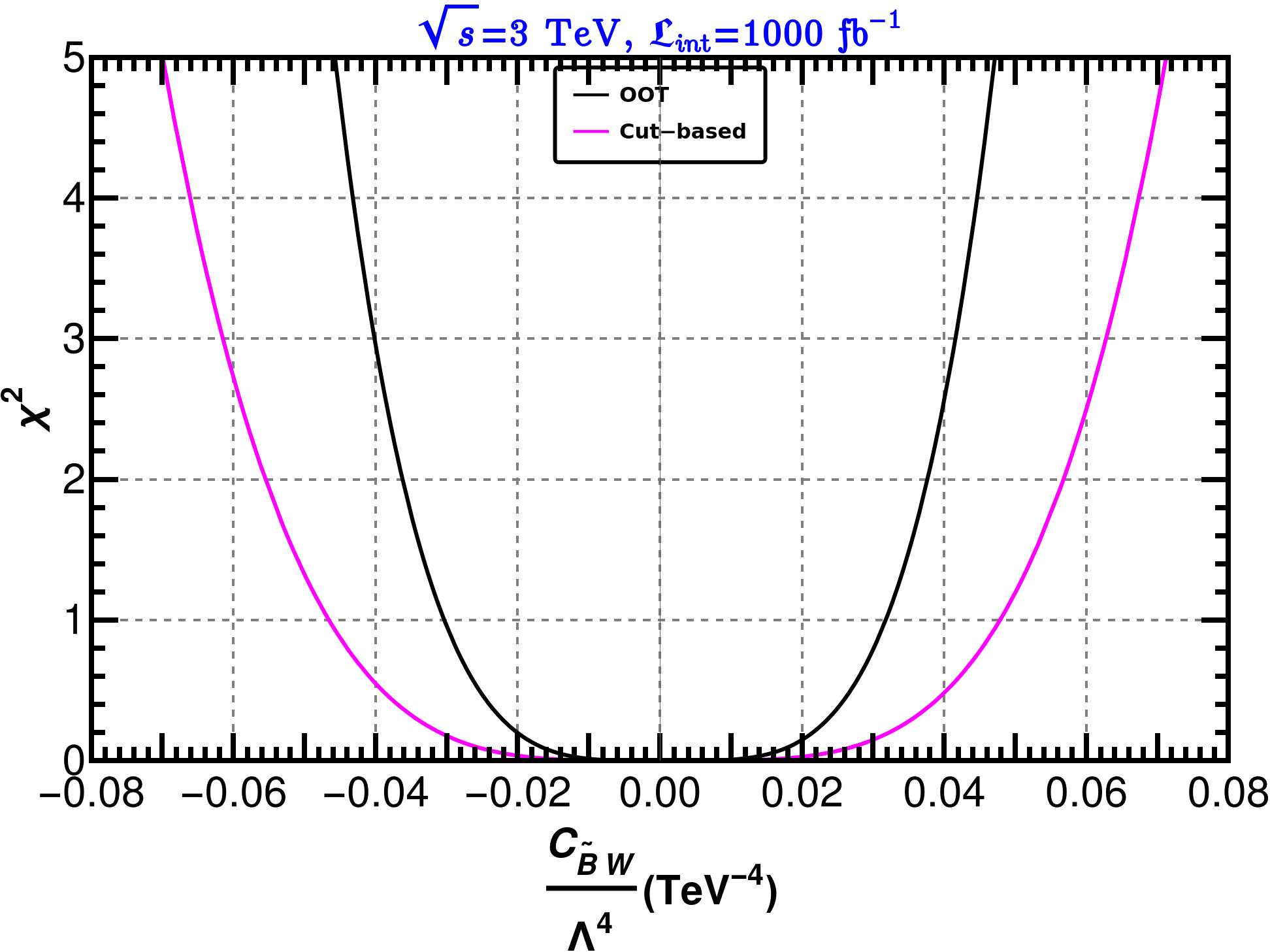}
	\includegraphics[height=5.5cm, width=7.2cm]{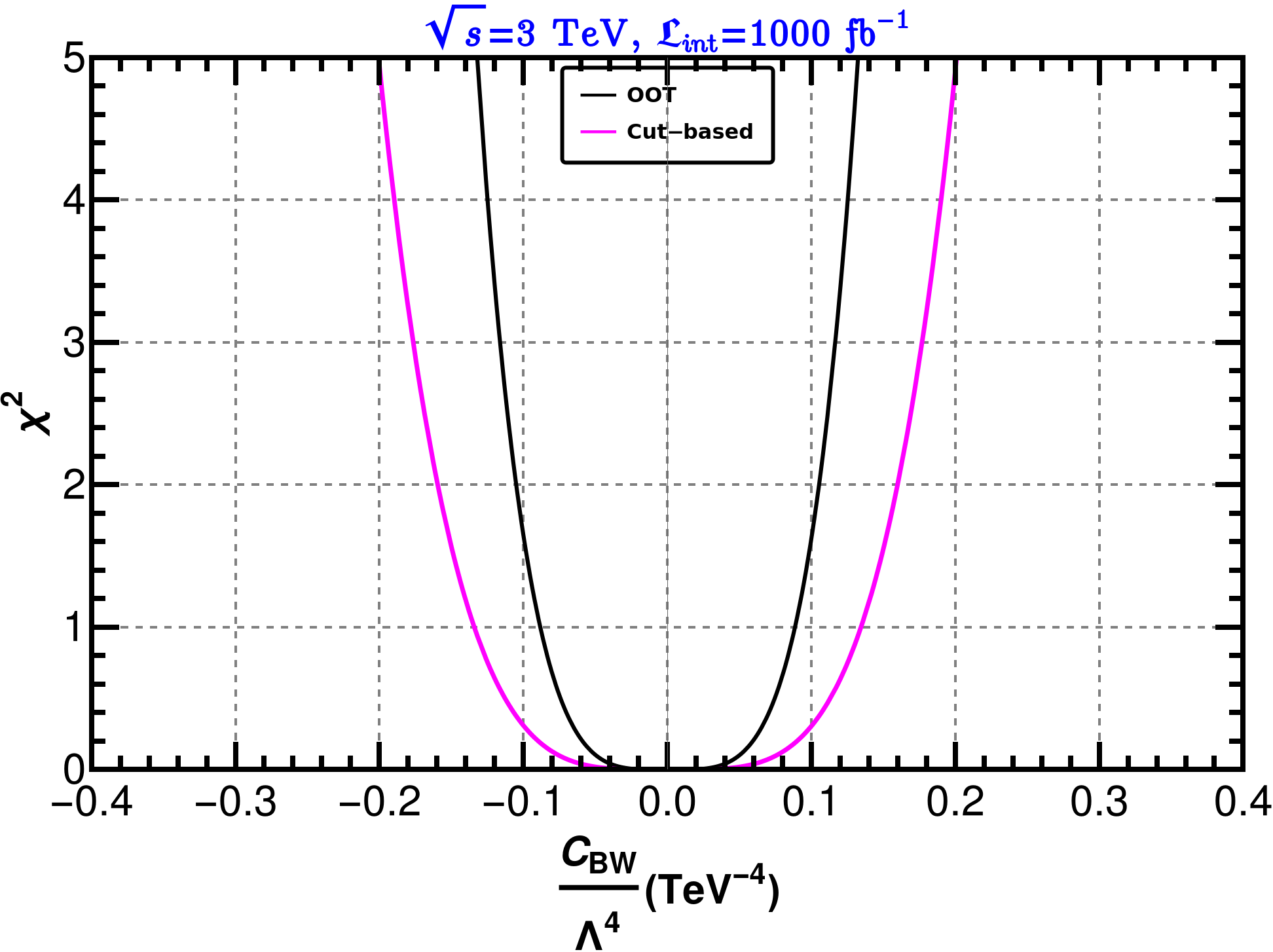}
	$$
	$$
	\includegraphics[height=5.5cm, width=7.2cm]{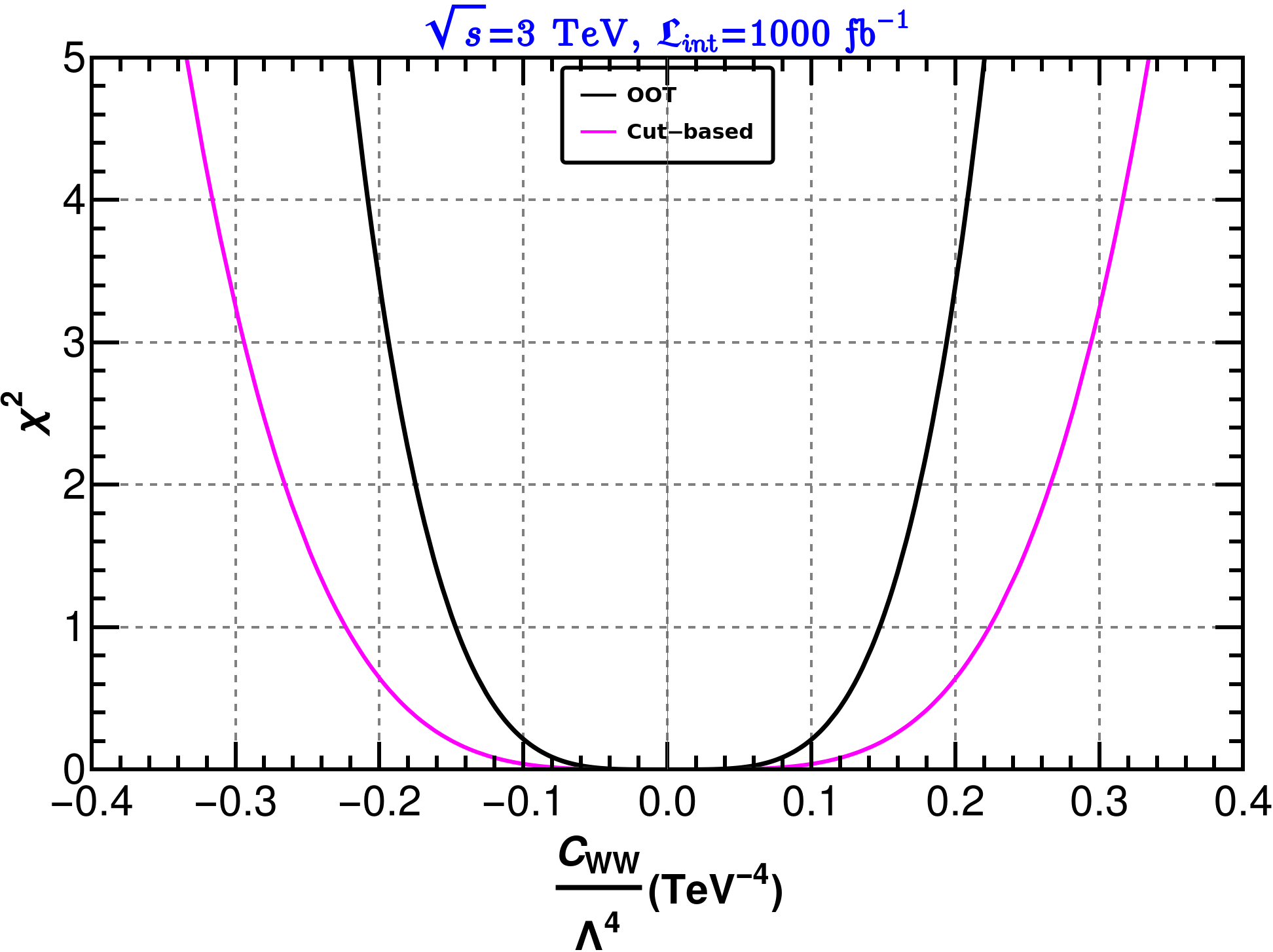}
	\includegraphics[height=5.5cm, width=7.2cm]{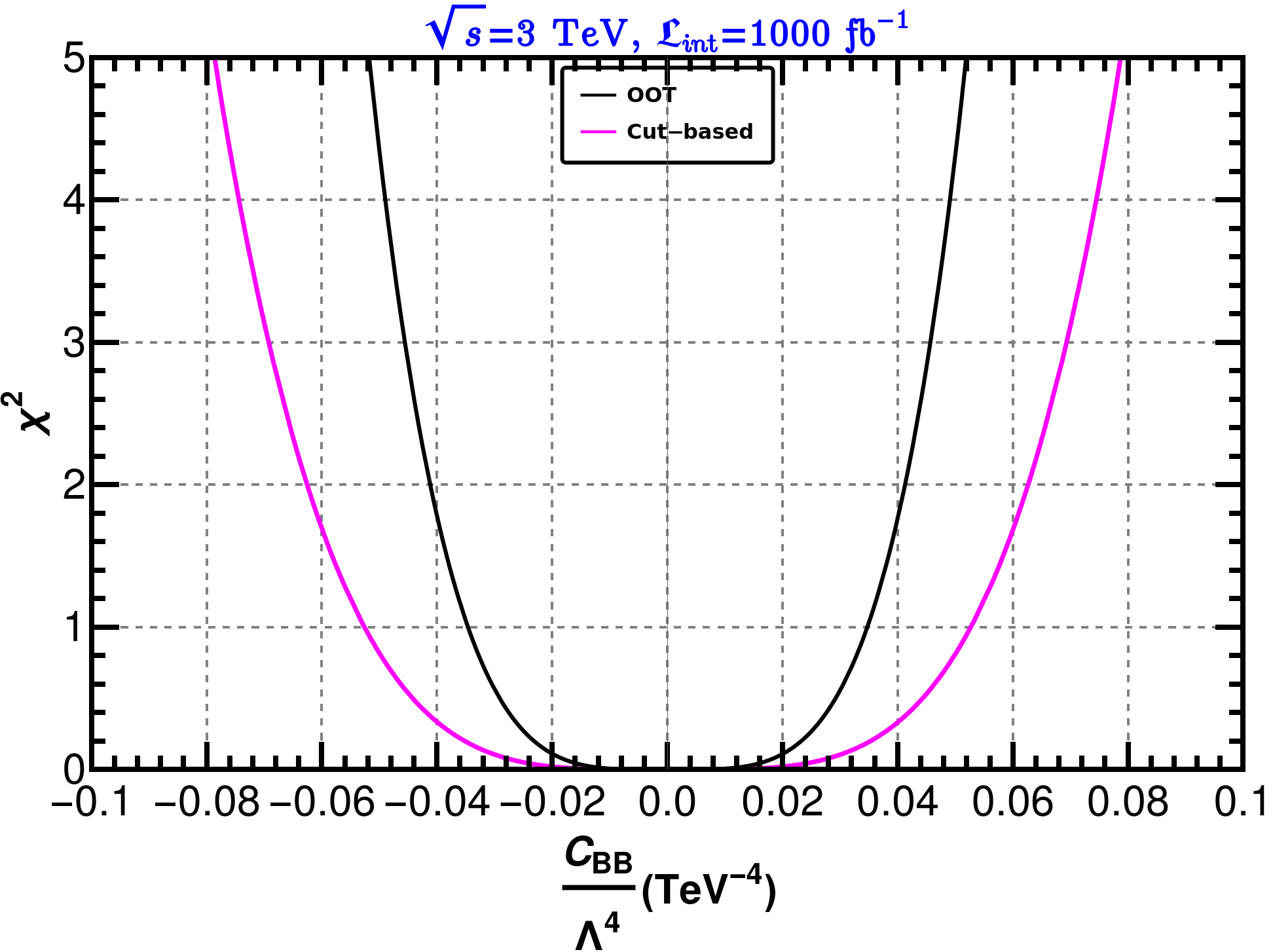}
	$$
	\caption{Comparison of the variation of the $\chi^2$ function of different NP couplings for unpolarized beam at the CLIC. Top left: $C_{\tilde B W}/\Lambda^4$; top right: $C_{ B W}/\Lambda^4$; bottom left: $C_{W W}/\Lambda^4$; bottom right: $C_{BB}/\Lambda^4$.} 	
	\label{fig:oot.col}
\end{figure}

\begin{table}
	\centering
	\begin{tabular}{| c | c |  c | c |  c | c | c | c   c | c  c | c  c| c  c| c c| } 
		\hline
		\multicolumn{1}{|c}{Couplings} &
		\multicolumn{3}{|c|}{$95\%$ C.L.} \\
		\cline{2-4}
		\multicolumn{1}{|c|}{$\rm (TeV^{-4})$}&
		\multicolumn{1}{c}{$P_{e^\pm} = ^{00\%}_{00\%}$}&
		\multicolumn{1}{|c}{$P_{e^\pm} = ^{+00\%}_{-80\%}$}&
		\multicolumn{1}{|c|}{$P_{e^\pm} = ^{+00\%}_{+80\%}$}\\
		\hline
		\multirow{2}*{$\frac{C_{\tilde{B}W}}{\Lambda^4}$}& $+0.044$ & $+0.050$  & $+0.038$  \\
		&  $-0.043$ & $-0.043$ & $-0.042$ \\
		\hline
		\multirow{2}*{$\frac{C_{BW}}{\Lambda^4}$} & $+0.125$  & $+0.299$ & $+0.085$ \\
		& $-0.125$ & $-0.299$ & $-0.085$ \\
		\hline
		\multirow{2}*{$\frac{C_{WW}}{\Lambda^4}$}& $+0.288$ &  $+0.307$ &  $+0.156$ \\
		& $-0.288$ & $-0.307$ & $-0.156$ \\
		\hline
		\multirow{2}*{$\frac{C_{BB}}{\Lambda^4}$}& $+0.049$ &  $+0.043$  & $+0.060$ \\
		& $-0.049$ & $-0.043$ & $-0.060$ \\
		\hline
	\end{tabular}
	\caption{Optimal statistical limit (95\% C.L.) on dim-8 nTGCs at CLIC for different beam polarization combinations.}
	\label{tab:95cl}
\end{table}

\begin{table}
	\centering
	\begin{tabular}{| c | c |  c | c |  c | c | c | c   c | c  c | c  c| c  c| c c| } 
		\hline
		\multicolumn{1}{|c}{Couplings} &
		\multicolumn{3}{|c|}{$95\%$ C.L.} \\
		\cline{2-4}
		\multicolumn{1}{|c|}{$\rm (TeV^{-4})$}&
		\multicolumn{1}{c}{$P_{e^\pm} = ^{00\%}_{00\%}$}&
		\multicolumn{1}{|c}{$P_{e^\pm} = ^{+00\%}_{-80\%}$}&
		\multicolumn{1}{|c|}{$P_{e^\pm} = ^{+00\%}_{+80\%}$}\\
		\hline
		\multirow{2}*{$\frac{C^{(2)}_{W^2B^2H^2}}{\Lambda^4}$}& $+0.121$ & $+0.121$ & $+0.118$\\
		                                                      & $-0.123$ & $-0.140$ & $-0.107$ \\
		\hline
		\multirow{2}*{$\frac{C^{(6)}_{WBH^2D^2}}{\Lambda^4}$} & $+0.088$ & $+0.100$ & $+0.076$\\
		                                                      & $-0.086$ & $-0.086$ & $-0.084$ \\
		\hline
		\multirow{2}*{$\frac{C^{(5)}_{WBH^2D^2}}{\Lambda^4}$} & $+0.086$ & $+0.086$ & $+0.084$ \\
		                                                      & $-0.088$ & $-0.100$ & $-0.076$ \\
		\hline
		\multirow{2}*{$\frac{C^{(2)}_{WBH^4}}{\Lambda^4}$}   &$+0.166$&$+0.166$&$+0.163$ \\
		                                                     & $-0.170$ & $-0.194$ & $-0.147$ \\
		\hline
		\multirow{2}*{$\frac{C^{(2)}_{WBH^2D^2}}{\Lambda^4}$}& $+0.022$ &  $+0.025$  & $+0.019$\\
		                                                     & $-0.021$ & $-0.021$ & $-0.042$\\
		\hline
		\multirow{2}*{$\frac{C^{(1)}_{W^2B^2H^2}}{\Lambda^4}$}& $+0.500$ & $+0.657$  & $+0.283$\\
		                                                      & $-0.500$ & $-0.657$ & $-0.283$ \\
		\hline
		\multirow{2}*{$\frac{C^{(4)}_{WBH^2D^2}}{\Lambda^4}$} & $+0.250$  & $+0.614$ & $+0.170$\\
		                                                      & $-0.250$ & $-0.614$ & $-0.170$ \\
		\hline
		\multirow{2}*{$\frac{C^{(3)}_{WBH^2D^2}}{\Lambda^4}$}& $+0.250$  & $+0.614$ & $+0.170$\\
		                                                     & $-0.250$  & $-0.614$ & $-0.170$ \\
		\hline
		\multirow{2}*{$\frac{C^{(1)}_{WBH^4}}{\Lambda^4}$}  & $+0.484$ &$+1.150$&$+0.329$ \\
		                                                    & $-0.484$ & $-1.150$ & $-0.329$ \\
		\hline
		\multirow{2}*{$\frac{C^{(1)}_{WBH^2D^2}}{\Lambda^4}$} & $+0.112$ & $+0.149$ & $+0.042$\\
		                                                    & $-0.112$  & $-0.149$ & $-0.042$\\
		\hline
		\multirow{2}*{$\frac{C^{(1)}_{W^3H^2}}{\Lambda^4}$} & $+0.434$  & $+0.463$ & $+0.235$ \\
		                                                    & $-0.434$ & $-0.463$ & $-0.235$ \\
		\hline
		\multirow{2}*{$\frac{C^{(5)}_{W^2H^2D^2}}{\Lambda^4}$}& $+0.144$&$+0.153$&$+0.078$ \\
		                                                      &$-0.144$&$-0.153$&$-0.078$ \\
		\hline
		\multirow{2}*{$\frac{C^{(1)}_{W^2H^4}}{\Lambda^4}$}&$+1.115$&$+1.189$&$+0.604$\\
		                                                   &$-1.115$&$-1.189$&$-0.604$\\
		\hline
		\multirow{2}*{$\frac{C^{(2)}_{W^2H^2D^2}}{\Lambda^4}$}& $+0.144$&$+0.153$&$+0.078$ \\
		                                                      & $-0.144$&$-0.153$&$-0.078$ \\
		\hline
		\multirow{2}*{$\frac{C^{(1)}_{B^2H^2D^2}}{\Lambda^4}$} & $+0.024$ & $+0.021$ &$+0.030$ \\
		                                                       & $-0.024$ & $-0.021$ &$-0.030$ \\
		\hline
		\multirow{2}*{$\frac{C^{(1)}_{B^2H^4}}{\Lambda^4}$}& $+0.190$ &$+0.166$&$+0.232$ \\
		                                                   & $-0.190$ &$-0.166$ &$-0.232$ \\
		\hline
		\multirow{2}*{$\frac{C^{(2)}_{B^2H^2D^2}}{\Lambda^4}$}& $+0.024$ & $+0.021$ &$+0.030$ \\
		                                                      & $-0.024$ & $-0.021$ &$-0.030$ \\
		\hline
	\end{tabular}
	\caption{Optimal statistical limit (95\% C.L.) on dim-8 effective couplings at CLIC for different beam polarization combinations related to bosonic operators using generally used operator basis.}
	\label{tab:95cl2}
\end{table}

\subsection{Sensitivity comparison: OOT vs cut-based analysis}
\label{sec:ootvscol}
The statistical limit of NP couplings can also be determined by a contemporary cut-based analysis. Here, we discuss the estimation of NP coupling through a cut-based analysis and compare the estimated limit with OOT results. The usual $\chi^2$ function for cut-based analysis is defined by
\begin{equation}
	\chi^2 =\sum^{\rm{bins}}_{j} \left(\frac{N_j^{\tt obs}-N_j^{\tt theo}(g_i)}{\Delta N_j}\right)^2,
	\label{eq:chi2.col}
\end{equation}
where $N_j^{\tt obs}$ and $N_j^{\tt theo}$ are the number of events from observation and theory in the $\rm{j^{\tt th}}$ bin of the differential cross-section distribution after employing all the cuts described in section~\ref{sec:coll}. The statistical uncertainty  in each bin is $\Delta N_j  = \sqrt{N^{\tt obs}_j}$, considering the number of events in each bin according to the Poisson distribution. Using Eq.~\eqref{eq:chi2.col}, the variation of the $\chi^2$ function with NP couplings  is shown in figure~\ref{fig:oot.col} by magenta color. The comparison plots clearly show that OOT provides a tighter bound on NP couplings than the cut-based analysis. If we compare the statistical limits, OOT performs better than the cut-based analysis by a factor of 1.7 for each NP coupling.  

\begin{figure}[htb!]
	\begin{align*}
		\includegraphics[height=5cm, width=4.9cm]{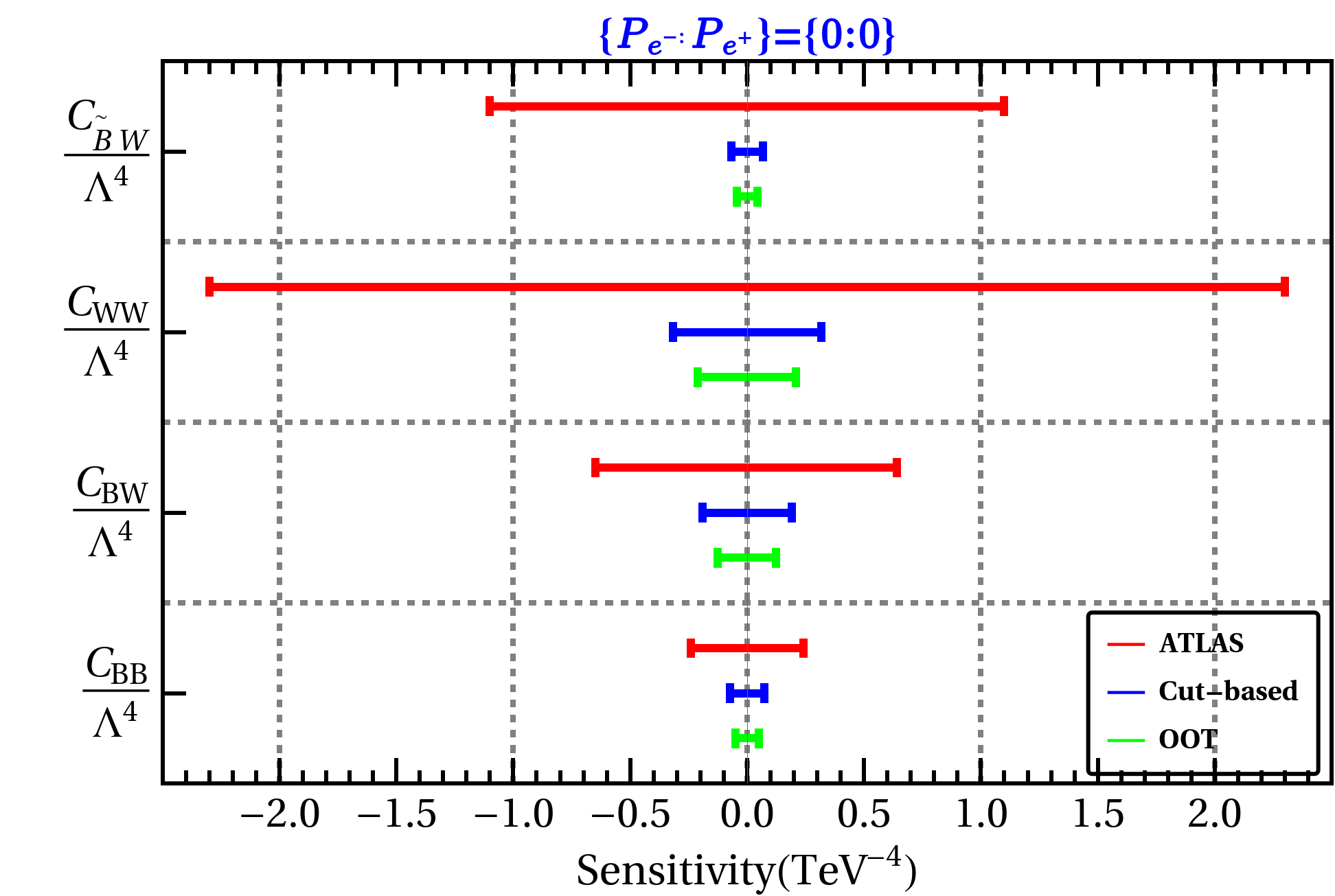} \quad
		\includegraphics[height=5cm, width=4.9cm]{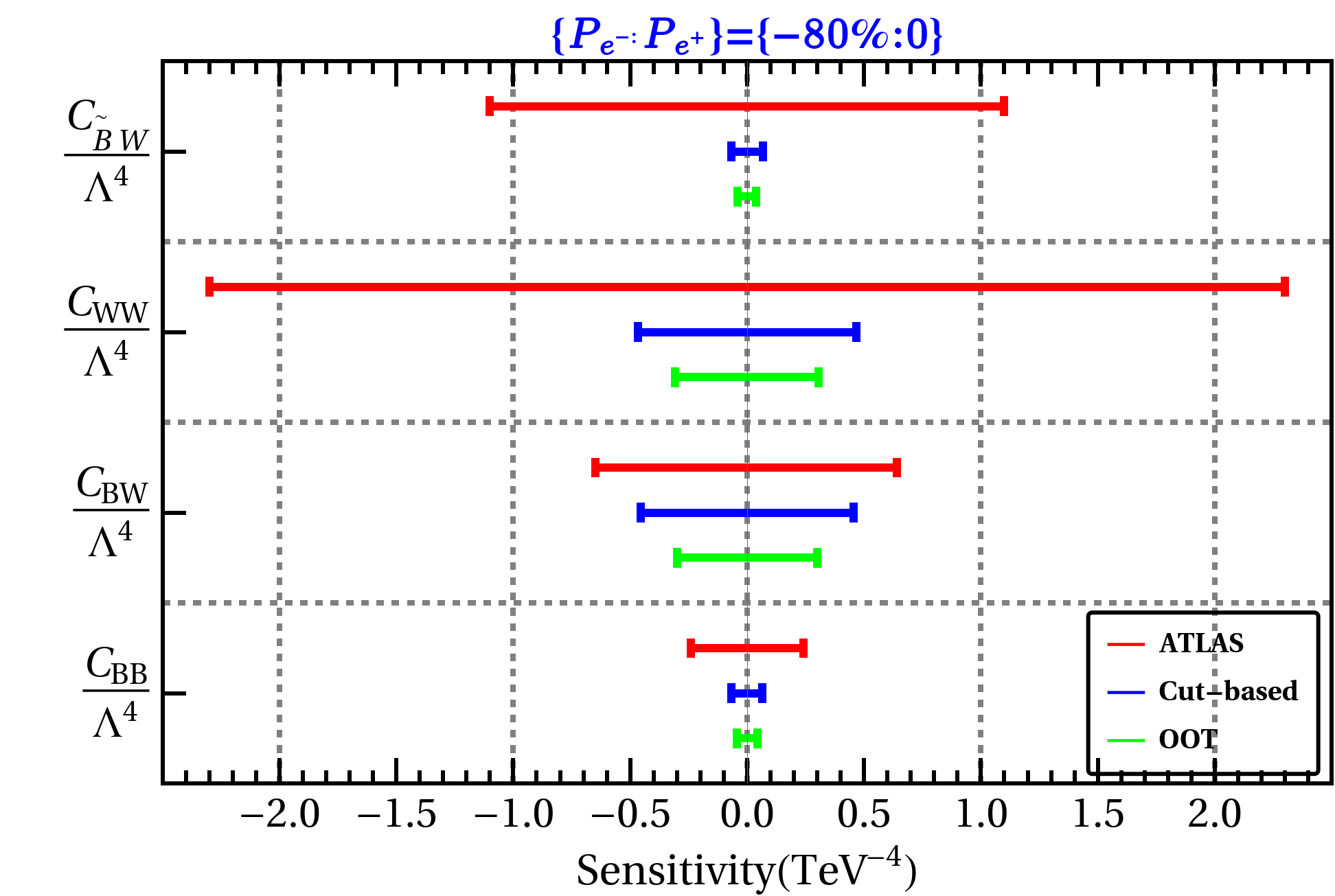} \quad
		\includegraphics[height=5cm, width=4.9cm]{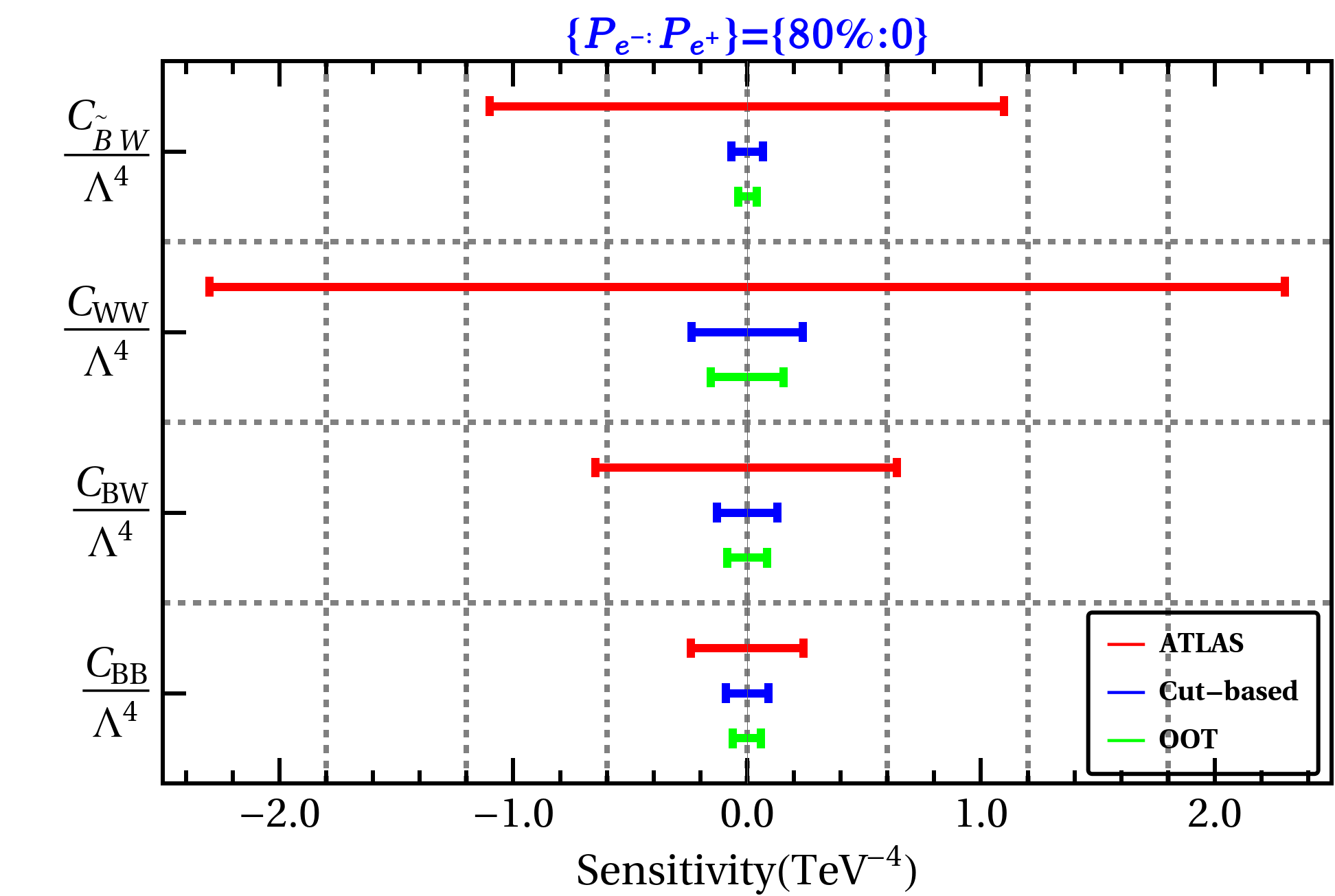} \quad
	\end{align*}
	\caption{Comparison of statistical limit (95\% C.L.) of NTGCs between latest ATLAS experiment, OOT and cut-based analysis. The information about beam polarization is written in the headings of the plots.} 	
	\label{fig:lim.comp}
\end{figure}
A comparison of 95\% C.L. of dim-8 couplings between the current experimental constrain from the ATLAS, the cut-based analysis and OOT is shown figure~\ref{fig:lim.comp} for various choices polarization combination where the benefit of the results using OOT can be clearly seen.
\subsection{Correlation between CP violating dim-8 nTGCs}
\label{sec:corelation}
The expected statistical limits at CLIC discussed above provide the tightest bound on each NP coupling in case of one parameter analysis. In a generic scenario, the estimation of a NP coupling can be affected by the presence of other NP couplings. In our case, all the CP-violating dim-8 NP couplings can contribute simultaneously to the process $e^+e^- \to ZZ$. Therefore, we consider two dim-8 nTGCs non-zero, at the time keeping the third one fixed at zero in order to estimate the statistical limits and correlation between these two.
\begin{figure}[htb!]
		\includegraphics[height=5cm, width=4.6cm]{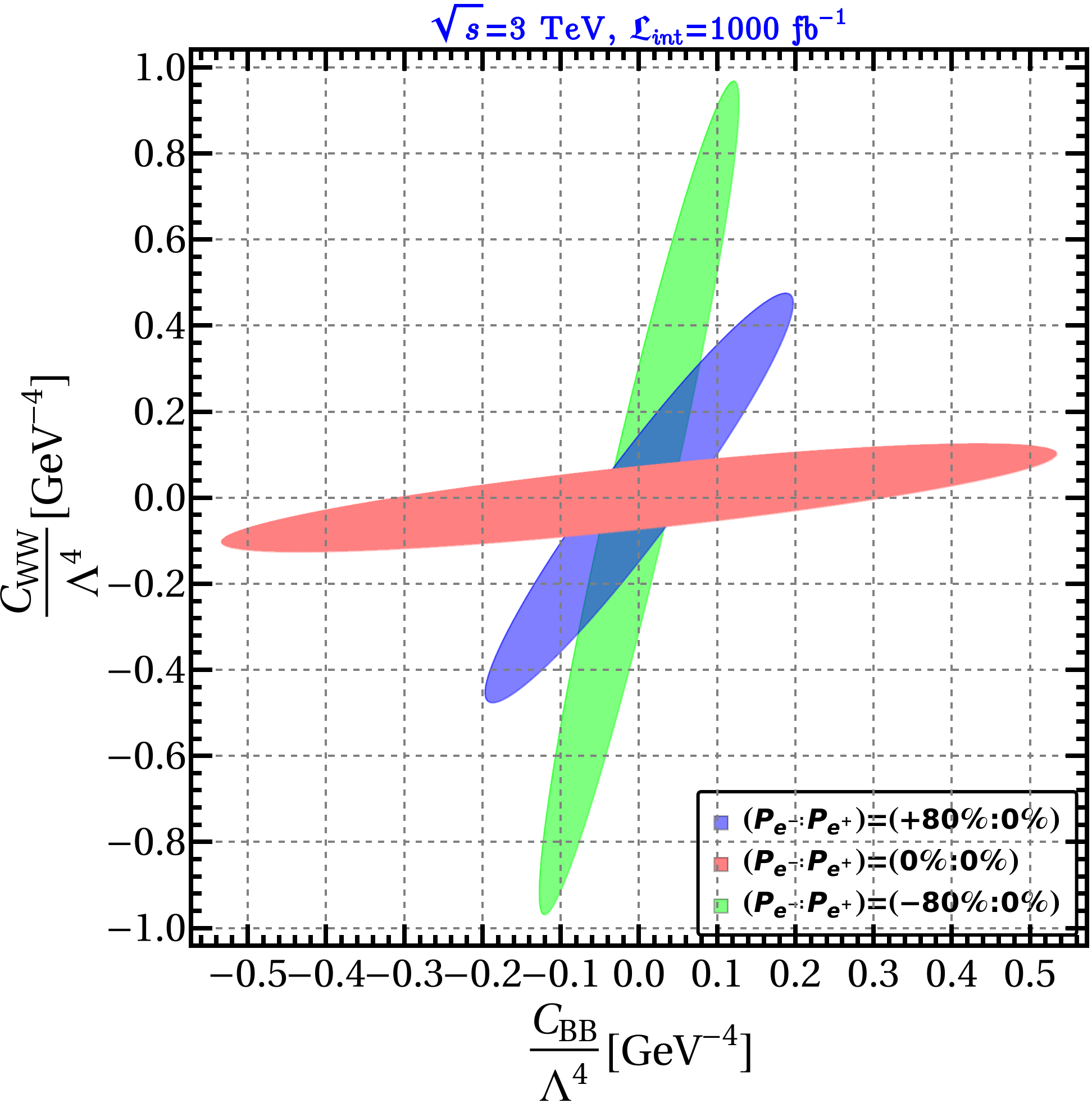} \quad
		\includegraphics[height=5cm, width=4.6cm]{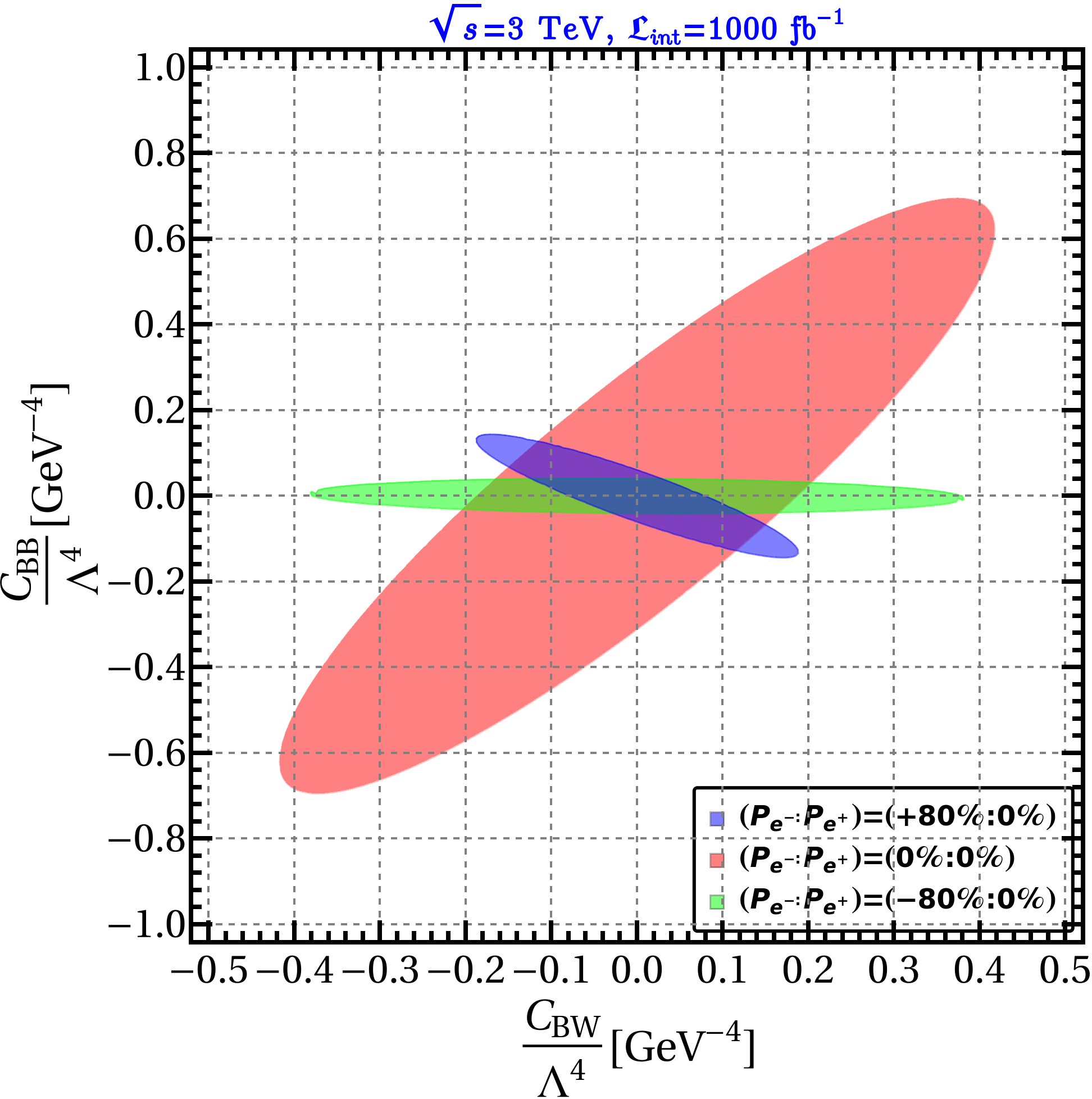} \quad
		\includegraphics[height=5cm, width=4.6cm]{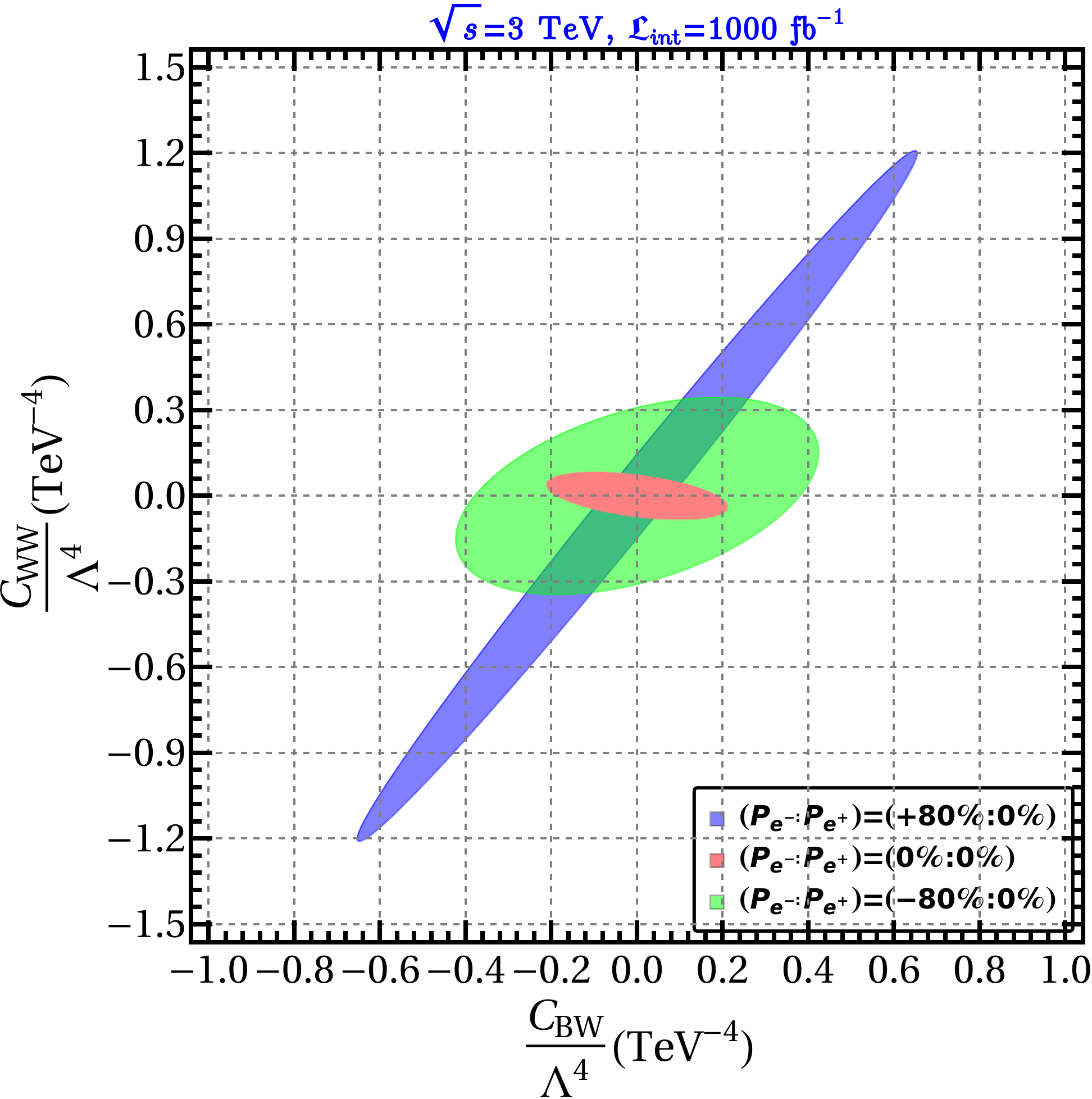} \quad
	\caption{Correlation among dim-8 nTGCs in 2 parameter space. Left:$\{C_{BB}/\Lambda^4,C_{WW}/\Lambda^4\}$, Middle:$\{C_{BW}/\Lambda^4,C_{BB}/\Lambda^4\}$, Right:$\{C_{BW}/\Lambda^4,C_{WW}/\Lambda^4\}$. Polarization information is written in the inset. CM energy and luminosity are written in the headings .}
	\label{fig:senscomp}
\end{figure}
In figure \ref{fig:senscomp}, we show 95\% C.L.  in a 2-parameter scenario spanned by planes $\{C_{BW}/\Lambda^4,C_{BB}/{\Lambda^4}$,
$\{C_{WW}/\Lambda^4,C_{BB}/\Lambda^4\}$ and $\{C_{BW}/\Lambda^4,C_{WW}/\Lambda^4\}$, respectively, for all polarization combination. It is worthwhile to notice that the optimal limit on dim-8 nTGCs in a 2-parameter case is still more stringent than experimental limit by CMS (one parameter analysis). Here, likewise one parameter scenario, the beam polarization plays a crucial roll to constraint a 2-parameter space. As expected, the estimated dim-8 couplings are abruptly changed due to the presence of others, depending on the constructive or destructive interference between themselves.

\section{Conclusion}
\label{sec:con}
In this paper, we have discussed the phenomenological aspects of exploring both CP-even and CP-odd dim-8 nTGCs couplings via an SMEFT formulation that offers an interesting window to look for BSM physics. We have considered the {\tt OSL} + missing energy final state for the $e^+e^- \to ZZ$ process to study the cut-based analysis for signal-background estimation. In this regard, kinematic cuts on missing energy ($E_{\tt miss}$), invariant di-lepton mass ($m_{\ell \ell}$) and $\Delta{R}_{\ell \ell}$ play a vital role to segregate the chosen signal from non-interfering SM backgrounds.

Next, we have adopted the optimal observable technique (OOT) to estimate the sensitivity of dim-8 nTCGs at the CLIC at CM energy $\sqrt{s}$ = 3 TeV with integrated luminosity $\mathfrak{L}_{\tt int}$ = 1000 ${\rm fb^{-1}}$ for different beam polarization combinations. The differential cross-section of the process $e^+e^- \to ZZ$ is taken as the observable to determine the optimal statistical limit of NP couplings in a sense that the covariance matrix is minimal. For a given CM energy and integrated luminosity, optimal 95\% C.L. on nTGCs is determined for all choices of the beam polarization combination. In comparison with the latest experimental bound on each NP coupling described above, the optimal sensitivity of NP couplings are 6 to 28 times better than existing ATLAS limit and 9 to 60 times better than CMS limits, depending on different dim-8 nTGCs.  We have shown the non-trivial correlations among CP-odd nTCGs. A generic cut-based analysis has also been performed in order to compare the sensitivity of NP couplings obtained through OOT at the same luminosity and CM energy. We have seen that in obtaining the statistical limit of NP couplings OOT performs significantly better than the cut-based analysis.

As far as we consider an $e^+e^-$ collider, this analysis illustrates that beam polarization plays an important role in constraining the statistical limit of dim-8 nTGCs more precisely. With an appropriate choice of beam polarization, the statistical limit of NP couplings is approximately $ 15\% ~ {\rm to} ~ 45\%$ more precise compared to the unpolarized beam. 

\acknowledgments

The author would like to thank Jayita Lahiri for useful suggestions and careful reading of the manuscript. SJ thanks Subhaditya Bhattacharya, Jose Wudka and Shakeel Ur Rahaman for useful discussions.

\appendix

\section{Expansion of dim-8 operators into a complete basis}
\label{sec:opsexpan}
In order to expand the operators listed in Eq.~\eqref{eq:dim8} into a complete basis, we consider the operator $\mathcal{O}_{WW}$ given by
\beq
\mathcal{O}_{W W}=H^{\dagger}W_{\mu\nu}W^{\mu\rho}\{D_{\rho},D^{\nu}\}H,
\label{eq:oww}
\eeq
with the anti-commutator given by
\beq
\{D_{\rho},D^{\nu}\}=[D_{\rho},D^{\nu}]+2D_{\rho}D^{\nu}, \qquad [D_{\rho},D^{\nu}] =-ig W_{\rho}^{\nu}-ig'B_{\rho}^{\nu}.
\eeq
Therefore, Eq.~\eqref{eq:oww} becomes
\begin{equation}
	\mathcal{O}_{WW}=-ig'(H^{\dagger}H)W_{\mu \nu} W^{\mu \rho} B^{\nu}_{\rho}-ig(H^{\dagger}H)W_{\mu \nu} W^{\mu \rho} W^{\nu}_{\rho}+2D^{\nu}(D_{\rho}H)H^{\dagger}W_{\mu \nu} W^{\mu \rho}.
	\label{eq:ops}
\end{equation}
After integration by parts, third term of Eq.~\eqref{eq:ops} turns into
\begin{align}
	\nonumber
	D^{\nu}(D_{\rho}H)H^{\dagger}W_{\mu \nu} W^{\mu \rho}&=(D_{\rho}H)(D^{\nu}H^{\dagger})W_{\mu \nu} W^{\mu \rho}+(D_{\nu}H)H^{\dagger}D^{\rho}(W_{\nu\mu} W^{\mu \rho})\\
	\label{eq:ibp1}
	&=(D_{\rho}H)(D^{\nu}H^{\dagger})W_{\mu \nu} W^{\mu \rho}+(D_{\nu}H)H^{\dagger}D^{\nu}(W_{\mu \rho} W^{\mu \rho}),\\\nonumber
\end{align}
which further simplifies to
\begin{equation}
	(D_{\nu}H)H^{\dagger}D^{\nu}(W_{\mu \rho} W^{\mu \rho})=(D^2H)H^{\dagger}(W_{\mu \rho} W^{\mu \rho})+(D^{\nu}H^{\dagger})(D_{\nu}H)(W_{\mu \rho} W^{\mu \rho}).
\end{equation}	
From the equation of motion of $H$, we get
\begin{equation}
	D^2H=\mu^2H-2\lambda (H^{\dagger}H)H-\text{fermion~densities}~(J^{\tt ff}), 
\end{equation}
where the expression of $J^{\tt ff}$ is given by
\begin{equation}
	J^{\tt ff} = y_u^{q} \bar{u}_R \sigma^2 q_L + y_d^{q}\bar{d}_R q_L + y^{e}\bar{e}_R l_L,
\end{equation}
with $\sigma^2= \left( \begin{array}{cc}
0 & -i \\
i & 0
\end{array} \right) $ is the Pauli matrix. Combining all these equations, we get
\begin{align}
	\nonumber
	\mathcal{O}_{WW} =& -ig'(H^{\dagger}H)W_{\mu \nu} W^{\mu \rho} B^{\nu}_{\rho}-ig(H^{\dagger}H)W_{\mu \nu} W^{\mu \rho} W^{\nu}_{\rho}+2(D_{\rho}H)(D^{\nu}H^{\dagger})W_{\mu \nu} W^{\mu \rho}\\\nonumber
	&-2\lambda(H^{\dagger}H)^2W_{\mu \rho} W^{\mu \rho}-2H^{\dagger}( y_u^{q}\bar{u}_R \sigma^2 q_L +  y_d^{q}\bar{d}_R q_L + y^{e} \bar{e}_R l_L)W_{\mu \rho} W^{\mu \rho}\\
	&+2(D^{\nu}H^{\dagger})(D_{\nu}H)(W_{\mu \rho} W^{\mu \rho}).\\\nonumber
\end{align}
If we re-express $\mathcal{O}_{WW}$ with the operators listed in \cite{Murphy:2020rsh}, then the relation becomes
\begin{align}
	\nonumber
	\mathcal{O}_{WW}=&-ig' Q^{(1)}_{W^2BH^2}-igQ^{(1)}_{W^3H^2}-2 \lambda Q^{(3)}_{W^2H^4}+2Q^{(1)}_{W^2H^2D^2}+2Q^{(2)}_{W^2H^2D^2}- y^q_u Q^{(1)}_{quW^2H}\\
	&- y^q_d Q^{(1)}_{qdW^2H}- y^e Q^{(1)}_{leW^2H}.
\end{align}
\bibliographystyle{JHEP}
\bibliography{ref-ntgc}
\end{document}